\documentclass[12pt,twoside]{article}

\usepackage{pslatex}	
\usepackage{amsfonts}
\usepackage{amsmath}
\usepackage{amsthm}
\usepackage{amscd}
\usepackage{cite}
\usepackage{epsf}
\usepackage{a4}

\def\beq{\begin{equation}}
\def\eeq{\end{equation}}
\def\bea{\begin{eqnarray}}
\def\eea{\end{eqnarray}}

\def\beann{\begin{eqnarray*}}
\def\eeann{\end{eqnarray*}}

\let\a=\alpha \let\be=\beta \let\g=\gamma \let\de=\delta
\let\e=\varepsilon \let\z=\zeta \let\h=\eta \let\th=\theta

 \let\k=\kappa \let\la=\lambda \let\m=\mu
\let\n=\nu \let\x=\xi \let\p=\pi \let\r=\rho \let\s=\sigma
\let\om=\omega 
\let\ph=\varphi  \let\PH=\Phi \let\Ps=\Psi
  \let\Th=\Theta
\let\La=\Lambda \let\G=\Gamma \let\D=\Delta

\let\qd=\quad  

\def\epp{\, .}
\def\epc{\, ,}

\def\tst#1{{\textstyle #1}}
\def\dst#1{{\displaystyle #1}}

\theoremstyle{plain}
\newtheorem{theorem}{Theorem}
\newtheorem{lemma}{Lemma}
\newtheorem{corollary}{Corollary}
\newtheorem*{corollary*}{Corollary}

\theoremstyle{definition}

\newtheorem*{remark}{Remark}
\newtheorem*{example}{Example}

\def\2{\frac{1}{2}} \def\4{\frac{1}{4}}

\def\6{\partial}

\def\+{\dagger}

\def\<{\langle} \def\>{\rangle}

\def\CO{{\cal O}}

\def\i{{\rm i}}

\def\re{{\rm e}}

\DeclareMathOperator{\sh}{sh}
\DeclareMathOperator{\ch}{ch}

 \def\End{{\rm End}} \def\id{{\rm id}}

\def\Re{{\rm Re\,}} \def\Im{{\rm Im\,}}

\def\res{{\rm res}}
\def\diag{{\rm diag}}

\def\tr{{\rm tr}}

\def\ev{\mathbf{e}}

\def\jv{\mathbf{j}}

\def\nv{\mathbf{n}}

\def\fa{\mathfrak{a}}
\def\faq{\overline{\mathfrak{a}}}
\def\fb{\mathfrak{b}}
\def\fbq{\overline{\mathfrak{b}}}

\renewcommand{\appendix}{%
   \renewcommand{\section}{
	\secdef\Appendix\sAppendix}%
   \setcounter{section}{0}%
   \renewcommand{\thesection}{\Alph{section}}%
   \renewcommand{\theequation}{\thesection.\arabic{equation}}%
}

\newcommand{\Appendix}[2][?]{%
     \refstepcounter{section}%
     \setcounter{equation}{0}%
     \addcontentsline{toc}{appendix}%
          {\protect\numberline{\appendixname~\thesection} #1}%
     \vspace{\baselineskip}%
     {\noindent\large\bfseries\appendixname: #2\par}%
     \sectionmark{#1}\vspace{\baselineskip}}

\newcommand{\sAppendix}[1]{%
     {\noindent\large\bfseries\appendixname\:: #1\par}%
     \sectionmark{#1}\vspace{\baselineskip}}

\begin{document}

\thispagestyle{empty}

\begin{center}

{\Large {\bf Integral representations for correlation functions
of the XXZ chain at finite temperature\\}}

\vspace{7mm}

{\large Frank G\"{o}hmann,\footnote[2]{e-mail:
goehmann@physik.uni-wuppertal.de} Andreas Kl\"{u}mper\footnote[1]{%
e-mail: kluemper@physik.uni-wuppertal.de} and Alexander Seel%
\footnote[3]{e-mail: seel@physik.uni-wuppertal.de}\\

\vspace{5mm}

Fachbereich Physik, Bergische Universit\"at Wuppertal,\\
42097 Wuppertal, Germany\\}

\vspace{20mm}

{\large {\bf Abstract}}

\end{center}

\begin{list}{}{\addtolength{\rightmargin}{10mm}
               \addtolength{\topsep}{-5mm}}
\item
We derive a novel multiple integral representation for a generating
function of the $\s^z$-$\s^z$ correlation functions of the spin-$\2$
XXZ chain at finite temperature and finite, longitudinal magnetic
field. Our work combines algebraic Bethe ansatz techniques for the
calculation of matrix elements with the quantum transfer matrix
approach to thermodynamics.
\\[2ex]
{\it PACS: 05.30.-d, 75.10.Pq}
\end{list}

\clearpage

\section{Introduction}
The quantum transfer matrix \cite{Suzuki85,SuIn87} is a means which
makes it possible to calculate finite temperature properties of
one-dimensional quantum spin systems. It becomes particularly efficient
\cite{Kluemper92,Kluemper93} for so-called integrable models which are
related to representations of the Yang-Baxter algebra and are solvable
by Bethe ansatz. Among these models the spin-$\2$ XXZ chain (henceforth
briefly referred to as `the' XXZ chain) is one of the most thoroughly
studied (see e.g.\ \cite{Gaudin83,Takahashi99,KBIBo}). In contrast to
many other integrable models, not only its spectral properties and
thermodynamics are well understood. In addition, efficient formulae for
norms \cite{Korepin82}, scalar products \cite{Slavnov89}, and the
multiple action of certain non-local operators \cite{KMST02a} are
available for the generalized model \cite{Korepin82} defined by the
$R$-matrix of the XXZ chain. These technical achievements led to the
derivation of certain multiple integral representations for (static)
two-point correlation functions and generating functions of two-point
functions at zero temperature \cite{KMST02a}.

In this work we show how the results of \cite{KMST02a} can be
generalized to finite temperatures. We concentrate on the generating
function of the $\s^z$-$\s^z$-correlation functions \cite{IzKo84}.
We use the quantum transfer matrix and the formalism of non-linear
integral equations in order to express it as a multiple integral over
the same auxiliary function that determines the free energy of the
model. In the zero temperature limit we recover the result of
\cite{KMST02a}. An extension of our results to other finite
temperature correlation functions of the XXZ chain is possible and
will be published elsewhere.

After providing the necessary foundations for treating correlation
functions at finite temperatures in section \ref{sec:found} we derive
our main result in section \ref{sec:result}, which is the integral
formula (\ref{mirep}) in theorem~\ref{theorem:main}. As a special case
of (\ref{mirep}) we obtain an integral representation for the so-called
emptiness formation probability \cite{KIEU94} of the XXZ chain at
finite temperature. In order to keep our presentation short we
formulate our results in the first instance only for the off-critical
regime ($\D > 1$) of the model. Note, however, that our method applies
for arbitrary values of the anisotropy parameter. In section
(\ref{sec:concl}) which is devoted to an outlook and to conclusions we
indicate how our results have to be modified in the critical regime.

\section{Foundations} \label{sec:found}
\subsection{\boldmath Hamiltonian and $R$-matrix}
\label{sec:rham}
The Hamiltonian of the $L$-site XXZ chain,
\begin{equation} \label{xxzham}
     H_{XXZ} = J \sum_{j=1}^L \Bigl( \s_{j-1}^x \s_j^x
                  + \s_{j-1}^y \s_j^y + \D (\s_{j-1}^z \s_j^z - 1)
		     \Bigr) \epc
\end{equation}
is defined on the $L$-fold tensor product $({\mathbb C}^2)^{\otimes L}$
through the elementary operators $\s_j^\a$, $\a = x, y, z$, which act
as Pauli matrices on the $j$th factor and trivially elsewhere. For
$j = 0$ we set $\s_0^\a = \s_L^\a$, thereby specifying the boundary
conditions to be periodic. $J > 0$ fixes the energy scale, and the
real parameter $\D$ is the anisotropy parameter of the model.

The Hamiltonian (\ref{xxzham}) is closely related to a certain
trigonometric solution
\begin{equation} \label{rxxz}
     R(\la,\m) = \begin{pmatrix}
                    1 & 0 & 0 & 0 \\
		    0 & \frac{\sh(\la - \m)}{\sh(\la - \m + \h)} &
		    \frac{\sh(\h)}{\sh(\la - \m + \h)} & 0 \\[1ex]
		    0 & \frac{\sh(\h)}{\sh(\la - \m + \h)} &
		    \frac{\sh(\la - \m)}{\sh(\la - \m + \h)}
		    & 0 \\
		    0 & 0 & 0 & 1
		 \end{pmatrix}
\end{equation}
of the Yang-Baxter equation which not only allows one to diagonalize
the Hamiltonian by means of the algebraic Bethe ansatz, but will
also enable us to associate a quantum transfer matrix with the model
that will be our principal tool for calculating its finite temperature
properties. We shall now briefly review the general theory.

For simplicity we shall restrict ourselves to fundamental models.
These are models which are completely determined by a numerical
$d^2 \times d^2$ matrix $R(\la,\m) \in \End \bigl( {\mathbb C}^d
\otimes {\mathbb C}^d \bigr)$ satisfying the Yang-Baxter equation%
\footnote{Here and in the following we sum over doubly occurring
Greek indices.}
\begin{equation} \label{ybe}
     R_{\a' \be'}^{\a \be} (\la, \m) R_{\a'' \g'}^{\a' \g} (\la, \n)
     R_{\be'' \g''}^{\be' \g'} (\m, \n) =
     R_{\be' \g'}^{\be \g} (\m, \n) R_{\a' \g''}^{\a \g'} (\la, \n)
     R_{\a'' \be''}^{\a' \be'} (\la, \m) \epp
\end{equation}
This equation is nowadays not usually presented in components, but
either graphically or in `operator form' with respect to the gl($d$)
standard basis\footnote{The basis consists of all $d \times d$ matrices
$e_\a^\be$ having a single 1 at the intersection of the $\a$th row with
the $\be$th column. Accordingly the basis matrices multiply as $e_\a^\be
e_\g^\de = \de^\be_\g e_\a^\de$. The canonical embedding is defined by
${e_j}_\a^\be = I_d^{\otimes (j-1)} \otimes e_\a^\be \otimes
I_d^{\otimes (L-j)}$ with $I_d$ being the $d \times d$ unit matrix.}
canonically embedded into $\bigl( {\mathbb C}^d \bigr)^{\otimes L}$.
In the latter case one defines
\begin{equation}
    R_{jk} (\la,\m) = R^{\a \g}_{\be \de} (\la,\m) {e_j}_\a^\be
                         {e_k}_\g^\de \epp
\end{equation}
Then (\ref{ybe}) is equivalent to
\begin{equation} \label{ybeop}
     R_{12} (\la,\m) R_{13} (\la,\n) R_{23} (\m,\n) =
        R_{23} (\m,\n) R_{13} (\la,\n) R_{12} (\la,\m) \epp
\end{equation}

A Hamiltonian associated with the $R$-matrix $R(\la,\m)$ is usually
constructed as follows: First of all an $L$-matrix at site $j$ with
matrix elements
\begin{equation} \label{deflmatrix}
     {L_j}^\a_\be (\la, \m) =
        R^{\a \g}_{\be \de} (\la, \m) {e_j}_\g^\de
\end{equation}
is introduced. These matrix elements are operators in $\bigl(
\End({\mathbb C}^d) \bigr)^{\otimes L}$. Multiplication of the
Yang-Baxter equation (\ref{ybe}) by ${e_j}_\g^{\g''}$ implies that
\begin{equation} \label{fundament}
     \check R(\la, \m) \big( L_j (\la, \n) \otimes L_j (\m, \n) \big) =
        \big( L_j (\m, \n) \otimes L_j (\la, \n) \big)
	\check R(\la, \m) \epc
\end{equation}
where $\check R^{\a \g}_{\be \de} = R^{\g \a}_{\be \de}$. One says that
$L_j (\la, \n)$ is a representation of the Yang-Baxter algebra with
$R$-matrix $R(\la,\m)$. This representation is called the fundamental
representation. Next, a monodromy matrix $T(\la)$ is introduced,
\begin{equation} \label{defmono}
     T(\la) = L_L (\la, \n) \dots L_1 (\la, \n) \epp
\end{equation}
Since $[{L_{j+1}}^\a_\be (\la, \n), {L_j}^\g_\de (\la, \n)] = 0$
it follows that
\begin{equation} \label{rtt}
     \check R(\la, \m) \big( T (\la) \otimes T (\m) \big) =
        \big( T (\m) \otimes T (\la) \big) \check R(\la, \m) \epp
\end{equation}
This means that also $T(\la)$ is a representation of the Yang-Baxter
algebra, a property which we use synonymously with integrability.

We shall assume that $R(\la,\m)$ is regular in the sense that $R(0,0) =
P$ is the permutation matrix acting on ${\mathbb C}^d \otimes
{\mathbb C}^d$. Then it follows that ${L_j}^\a_\be (0,0) =
{e_j}^\a_\be$ and further, setting $\n = 0$ in (\ref{defmono}), that
the transfer matrix, defined as
\begin{equation}
     t(\la) = \tr ( T(\la)) \epc
\end{equation}
has the expansion
\begin{equation} \label{fundhsum}
     t(\la) = \hat U \exp \bigl\{ \la H + \CO (\la^2) \bigr\}
\end{equation}
around $\la = 0$. Here $\hat U = t(0)$ is the shift operator and
\begin{equation} \label{fundhdens}
     H = \hat U^{-1} t'(0) = \sum_{j=1}^L
            \6_\la \check R_{j-1, j} (\la,0) \Big|_{\la = 0}
\end{equation}
is the Hamiltonian associated with the fundamental representation
(\ref{deflmatrix}) of the Yang-Baxter algebra. Note that periodic
boundary conditions, $\check R_{01} = \check R_{L1}$, are implied in%
~(\ref{fundhdens}).

Employing the usual conventions for tensor products we may interpret
the matrix (\ref{rxxz}) as a linear operator in ${\mathbb C}^2 \otimes
{\mathbb C}^2$. It is then easy to show that it satisfies the
Yang-Baxter equation (\ref{ybe}). Inserting (\ref{rxxz}) into the
right hand side of (\ref{fundhdens}) we obtain
\begin{equation} \label{xxzham2}
     H = \frac{1}{2 \sh(\h)} \sum_{j=1}^L \Bigl( \s_{j-1}^x \s_j^x
                  + \s_{j-1}^y \s_j^y + \ch(\h) (\s_{j-1}^z \s_j^z - 1)
		     \Bigr) \epp 
\end{equation}
Thus,
\begin{equation} \label{rescaleham}
     H_{XXZ} = 2J \sh(\h) H
\end{equation}
if we identify $\D = \ch(\h)$.

\subsection{The Trotter-Suzuki formula}
\label{sec:trosuz}
Consider the elementary formula
\begin{equation} \label{euler2}
     \lim_{N \rightarrow \infty} \Bigl( 1 + \tst{\frac{X_N}{N}} \Bigr)^N
        = \re^X \epc
\end{equation}
valid for any sequence $(X_N)_{N \in {\mathbb N}}$ of complex numbers
converging to $X$. It still works \cite{Suzuki85} for a matrix (an
operator) $X = - \be H$. Hence, (\ref{fundhsum}) and (\ref{euler2})
imply that
\begin{equation} \label{stform}
     \lim_{N \rightarrow \infty} \Bigl( \hat U^{-1}
        t \bigl( - \tst{\frac{\be}{N}} \bigr) \Bigr)^N =
     \lim_{N \rightarrow \infty} \Bigl( 1 + \tst{\frac{1}{N}}
        \bigl( - \be H  + \CO \bigl( \tst{\frac{1}{N}} \bigr) \bigr)
	  \Bigr)^N
	= \re^{- \be H} \epp
\end{equation}
This is a special form of the Trotter-Suzuki formula for the statistical
operator $\re^{- H/T}$ if we choose $\be$ to be the inverse
temperature, $\be =  1/T$.

\subsection{The quantum transfer matrix}
\label{sec:qtm}
We would like to use the Trotter-Suzuki formula (\ref{stform}) in a way
that is compatible with the special structure of integrable models.
In this context it will turn out to be useful to introduce $N$ auxiliary
spaces denoted by indices $\bar 1, \dots, \bar N$ and to work with
$R$-operators rather than $L$-matrices.

We introduce two types of monodromy matrices: type one,
\begin{equation}
     T_{\bar j} (\la) = R_{\bar j L} (\la,\m) \dots
                           R_{\bar j  1} (\la,\m)  \epc
\end{equation}
and type two,
\begin{equation}
     \overline T_{\bar j} (\la) = R_{1 \bar j} (\m,\la) \dots
                                R_{L \bar j} (\m,\la) \epp
\end{equation}
\begin{remark}
If $R$ is unitary, $R_{12} (\la,\m) R_{21} (\m,\la) = \id$, then
$\overline T_{\bar j} (\la) = T^{- 1}_{\bar j} (\la)$.
\end{remark}

By construction the trace of the monodromy matrix $T_{\bar j} (\la)$
with respect to the auxiliary space  is the usual transfer matrix,
$t(\la) = \tr_{\bar j} (T_{\bar j} (\la))$. The structure of
$\bar t(\la) = \tr_{\bar j} (\overline T_{\bar j} (\la))$ can most
easily be explored by means of the parity operator $\hat P$ which we
define by its action on the canonical basis of local operators,
$\hat P {e_j}^\a_\be \hat P = {e_{L - j + 1}}^\a_\be$. Then
\begin{equation} \label{parmonobar}
     \overline T_{\bar j} (\la) = \hat P (PRP)_{\bar j L} (\m,\la)
        \dots (PRP)_{\bar j 1} (\m,\la) \hat P \epp
\end{equation}
Here one has to be careful not to confuse the parity operator
$\hat P$ with the permutation matrix $P$. In (\ref{parmonobar}) we set
$\m = 0$, take the trace in space $\bar j$ and take into account that
the parity operator acts only in quantum space. We infer with
(\ref{fundhsum}), (\ref{fundhdens}) that
\begin{align} \label{tbarexp} \notag
     \bar t(\la) & = \hat P \hat U \exp \bigl\{ \la \sum_{j=1}^L
                        \6_\la (PRP\check)_{j-1 j} (0,\la)
			\big|_{\la = 0} + \dots \bigr\} \hat P\\[-1ex]
                 & = \exp \bigl\{ \la \sum_{j=1}^L 
                        \6_\la \check R_{j-1 j} (0,\la)
			\big|_{\la = 0} + \CO (\la^2) \bigr\}
			\hat U^{- 1} \epc
\end{align}
where we used in the second equation that $(PRP)_{j-1 j} =
R_{j j-1}$ and $\hat P \hat U \hat P = \hat U^{- 1}$.

In the following we shall assume that
\begin{equation} \label{hamtbar}
     \6_\la \check R_{j-1 j} (0,\la) \Big|_{\la = 0} =
            - \6_\la \check R_{j-1, j} (\la,0) \Big|_{\la = 0} \epp
\end{equation}
This is certainly true for $R$-matrices of difference form like
(\ref{rxxz}) but also for other more general $R$-matrices like
the $R$-matrix of the Hubbard model. Combining (\ref{fundhsum}),
(\ref{fundhdens}), (\ref{tbarexp}) and (\ref{hamtbar}) we obtain
\begin{equation}
     \r_{N,L} := \Bigl( \bar t \bigl( \tst{\frac{\be}{N}} \bigr)
                     t \bigl( - \tst{\frac{\be}{N}} \bigr)
		     \Bigr)^\frac{N}{2}
	       = \Bigl( 1 + \tst{\frac{2}{N}}
	         \Bigl( - \be H 
		  + \CO \bigl( \tst{\frac{1}{N}} \bigr) \Bigr)
		  \Bigr)^\frac{N}{2} \epp
\end{equation}
Using the second equation (\ref{stform}) we conclude that
\begin{equation} \label{limznl}
     \lim_{N \rightarrow \infty} \r_{N,L} = \re^{- \be H} \epp
\end{equation}

The latter expression is very useful for integrable models. It leads
us rather directly to the notion of the quantum transfer matrix, for
we may write
\begin{align} \label{piovertwo}
      \r_{N,L}
        & = \tr_{\bar 1 \dots \bar N} \bigl\{
	     \overline T_{\bar N} \bigl( \tst{\frac{\be}{N}} \bigr)
	     T_{\overline{N-1}} \bigl( - \tst{\frac{\be}{N}} \bigr)
	     \dots 
	     \overline T_{\bar 2} \bigl( \tst{\frac{\be}{N}} \bigr)
	     T_{\bar 1} \bigl( - \tst{\frac{\be}{N}} \bigr)
	     \bigr\} \notag \\[.5ex]
        & = \tr_{\bar 1 \dots \bar N} \bigl\{
	     \overline T_{\bar N} \bigl( \tst{\frac{\be}{N}} \bigr)
	     T_{\overline{N-1}}^t \bigl( - \tst{\frac{\be}{N}} \bigr)
	     \dots 
	     \overline T_{\bar 2} \bigl( \tst{\frac{\be}{N}} \bigr)
	     T_{\bar 1}^t \bigl( - \tst{\frac{\be}{N}} \bigr)
	     \bigr\} \notag \\[.5ex]
        & = \tr_{\bar 1 \dots \bar N} \bigl\{
	     R_{1 \bar N} \bigl(\m, \tst{\frac{\be}{N}} \bigr)
	     R_{2 \bar N} \bigl(\m, \tst{\frac{\be}{N}} \bigr) \dots
	     R_{L \bar N} \bigl(\m, \tst{\frac{\be}{N}} \bigr)
	     \notag \\
          & \mspace{72mu} R_{\overline{N-1} \, 1}^{t_1}
	     \bigl(- \tst{\frac{\be}{N}}, \m \bigr)
             R_{\overline{N-1} \, 2}^{t_1}
	     \bigl(- \tst{\frac{\be}{N}}, \m \bigr) \dots
             R_{\overline{N-1} \, L}^{t_1}
	     \bigl(- \tst{\frac{\be}{N}}, \m \bigr)
	     \notag \\ & \mspace{180mu} \dots
	     \notag \\
          & \mspace{72mu}
	     R_{1 \bar 2} \bigl(\m, \tst{\frac{\be}{N}} \bigr)
	     R_{2 \bar 2} \bigl(\m, \tst{\frac{\be}{N}} \bigr) \dots
	     R_{L \bar 2} \bigl(\m, \tst{\frac{\be}{N}} \bigr)
	     \notag \\
          & \mspace{72mu} R_{\bar 1 1}^{t_1}
	     \bigl(- \tst{\frac{\be}{N}}, \m \bigr)
             R_{\bar 1 2}^{t_1}
	     \bigl(- \tst{\frac{\be}{N}}, \m \bigr) \dots
             R_{\bar 1 L}^{t_1}
	     \bigl(- \tst{\frac{\be}{N}}, \m \bigr)
             \bigr\} \Bigr|_{\m = 0} \notag \\[.5ex]
        & = \tr_{\bar 1 \dots \bar N} \bigl\{
	     R_{1 \bar N} \bigl(\m, \tst{\frac{\be}{N}} \bigr)
             R_{\overline{N-1} \, 1}^{t_1}
	     \bigl(- \tst{\frac{\be}{N}}, \m \bigr) \dots
             R_{\bar 1 1}^{t_1} \bigl(- \tst{\frac{\be}{N}}, \m \bigr)
	     \notag \\ & \mspace{180mu} \dots
	     \notag \\
          & \mspace{72mu}
	     R_{L \bar N} \bigl(\m, \tst{\frac{\be}{N}} \bigr)
             R_{\overline{N-1} \, L}^{t_1}
	     \bigl(- \tst{\frac{\be}{N}}, \m \bigr) \dots
             R_{\bar 1 L}^{t_1}
	     \bigl(- \tst{\frac{\be}{N}}, \m \bigr)
             \bigr\} \Bigr|_{\m = 0} \notag \\[.5ex]
        & = \tr_{\bar 1 \dots \bar N} \bigl\{
	     T^{QTM}_1 (0) \dots T^{QTM}_L (0) \bigr\} \epp
\end{align}
Here the transpose with respect to space 1 occurring in the third
equation is defined by ${R^{t_1}}^{\a \g}_{\be \de} =
R^{\be \g}_{\a \de}$. Notice that we have reordered the product of
$R$-matrices in the fourth equation and have introduced another
monodromy matrix,
\begin{equation} \label{monoqtm}
     T^{QTM}_j (\la ) =
        R_{j \bar N} \bigl(\la, \tst{\frac{\be}{N}} \bigr)
	R_{\overline{N-1} \, j}^{t_1}
	   \bigl(- \tst{\frac{\be}{N}}, \la \bigr) \dots
        R_{j \bar 2} \bigl(\la, \tst{\frac{\be}{N}} \bigr)
	R_{\bar 1 j}^{t_1} \bigl(- \tst{\frac{\be}{N}}, \la \bigr)
\end{equation}
in the fifth equation. This monodromy matrix and the corresponding
transfer matrix
\begin{equation}
     t^{QTM} (\la ) = \tr_j T^{QTM}_j (\la ) \epc
\end{equation}
which will be called the quantum transfer matrix, are our main tools
for exploring the finite temperature properties of the XXZ chain below.

The quantum transfer matrix is useful mainly for two reasons. First,
in the thermodynamic limit $L \rightarrow \infty$ a single leading
eigenvalue of the quantum transfer matrix determines the  free energy.
Second, the spectrum of the quantum transfer matrix at finite Trotter
number $N$ can be calculated by means of the algebraic Bethe ansatz.
Let us explain this in more detail:

Using (\ref{piovertwo}) the partition function for a quantum chain of
length $L$ is calculated as
\begin{equation} \label{finitepart}
     Z_L = \lim_{N \rightarrow \infty} \:
           \underbrace{\tr_{1 \dots L} \, \r_{N,L}}_{=: Z_{N,L}}
         = \lim_{N \rightarrow \infty} \tr_{\bar 1 \dots \bar N}
	      \bigl( t^{QTM} (0) \bigr)^L
         = \sum_{n=1}^\infty \La^L_n (0) \epc
\end{equation}
where the $\La_n (0)$ are the eigenvalues of the quantum transfer
matrix in the Trotter limit $N \rightarrow \infty$ at spectral
parameter $\la = 0$. The free energy per lattice site in the
thermodynamic limit follows as
\begin{equation}
     f = - \lim_{L \rightarrow \infty} \lim_{N \rightarrow \infty}
            \frac{\ln(Z_{N,L})}{\be L}
       = - \frac{\ln(\La_0 (0))}{\be} 
\end{equation}
if $\La_0 (0)$ is the (finitely degenerate) leading eigenvalue of the
quantum transfer matrix. For the remainder we shall accept the
following.
\begin{enumerate}
\item
The limits $L \rightarrow \infty$ and $N \rightarrow \infty$ are
interchangeable \cite{Suzuki85,SuIn87}.
\item
The leading eigenvalue of the quantum transfer matrix is non-degenerate,
real and positive and in the Trotter limit is separated from the
next-to-leading eigenvalues by a gap.
\end{enumerate}
The thermodynamics of the model under scrutiny is then determined
by the leading eigenvalue $\La_0 (0)$ of the quantum transfer matrix
in the Trotter limit $N \rightarrow \infty$.

We have seen that instead of calculating the infinitely many
eigenvalues of the usual transfer matrix which determine the
spectrum of the Hamiltonian it suffices to calculate a single
eigenvalue of the quantum transfer matrix. Now we shall see that
the quantum transfer matrix is a representation of the same Yang-Baxter
algebra as the ordinary transfer matrix. Taking the transpose with
respect to space 1 in (\ref{ybeop}) we obtain
\begin{equation} \label{ybeopt}
     R_{23} (\la,\m) R_{12}^{t_1} (\n,\la) R_{13}^{t_1} (\n,\m) =
        R_{13}^{t_1} (\n,\m) R_{12}^{t_1} (\n,\la) R_{23} (\la,\m) \epp
\end{equation}
Then, because of (\ref{ybeop}) and (\ref{ybeopt}), the monodromy matrix
(\ref{monoqtm}) provides a representation of the Yang-Baxter algebra,
\begin{equation} \label{ybaop}
     R_{jk} (\la,\m) T^{QTM}_j (\la) T^{QTM}_k (\m) =
          T^{QTM}_k (\m) T^{QTM}_j (\la) R_{jk} (\la,\m) \epp
\end{equation}
This means that the leading eigenvalue of the quantum transfer matrix
may be obtained by algebraic Bethe ansatz (or possibly by one of the
other powerful methods connected with the Yang-Baxter algebra as e.g.\
the method of separation of variables \cite{Sklyanin95}).

\subsection{Correlation functions within the quantum transfer matrix
approach} \label{sec:corqtm}
The quantum transfer matrix approach allows us to calculate, in
principle, finite temperature correlation functions of local operators
by means of the algebraic Bethe ansatz. Here we shall consider
correlation functions of the form
\begin{equation} \label{defcorloc}
     \bigl\< X_j^{(1)} \dots X_k^{(k-j+1)} \bigr\>_T =
        \lim_{L \rightarrow \infty}
	\frac{\tr_{1, \dots, L} \, \re^{- \be H} 
                                        X_j^{(1)} \dots X_k^{(k-j+1)}}
	     {\tr_{1, \dots, L} \, \re^{- \be H}} \epc
\end{equation}
where $j, k \in \{1, \dots, L\}$, $j \le k$ and the $X_m^{(n)}$ are
arbitrary local operators. Using (\ref{limznl}) and (\ref{piovertwo})
we rewrite (\ref{defcorloc}) as
\begin{align} \label{corloc}
     \bigl\< X_j^{(1)} & \dots X_k^{(k-j+1)} \bigr\>_T \notag \\
        = & \lim_{N, L \rightarrow \infty}
        \tr_{\bar 1 \dots \bar N} \, \tr_{1 \dots L} \,
	T^{QTM}_1 (0) \dots T^{QTM}_L (0)
	X_j^{(1)} \dots X_k^{(k-j+1)} \big/ Z_{N,L} \notag \\
        = & \lim_{N, L \rightarrow \infty}
        \tr_{\bar 1 \dots \bar N} \, \bigl( t^{QTM} (0) \bigr)^{j-1}
	\tr \{T^{QTM} (0) X^{(1)}\} \dots
	\tr \{T^{QTM} (0) X^{(k-j+1)}\} \notag \\[-1ex]
	& \mspace{360mu} \cdot \bigl( t^{QTM} (0) \bigr)^{L-k}
	\big/ Z_{N,L} \notag \\[.5ex]
        = & \lim_{N, L \rightarrow \infty}
	  \frac{\sum_{n=0}^{d^N - 1} \La_n^{L - k + j - 1} (0)
	        \<\Ps_n| \tr \{T^{QTM} (0) X^{(1)}\} \dots
		         \tr \{T^{QTM} (0) X^{(k-j+1)}\} |\Ps_n\>}
	       {\sum_{n=0}^{d^N - 1} \La_n^L (0)} \notag \\[.5ex]
        = & \lim_{N \rightarrow \infty} \La_0^{j - k - 1} (0)
	        \<\Ps_0| \tr \{T^{QTM} (0) X^{(1)}\} \dots
		         \tr \{T^{QTM} (0) X^{(k-j+1)}\} |\Ps_0\> \epp
\end{align}
Here we assumed that the quantum transfer matrix is similar to a
diagonal matrix\footnote{This is known for the XXZ chain, but not in
the general case.} and has `normalized' eigenvectors $|\Ps_n\>$. We
see that a single normalized eigenvector $|\Ps_0\>$ and the
corresponding eigenvalue $\La_0 (\la)$ of the quantum transfer matrix
completely determine the finite temperature correlation functions of
local operators.

Let us consider two examples related to the XXZ chain. In this case
$d = 2$, and we can represent the monodromy matrix (\ref{monoqtm}) as a
$2 \times 2$ matrix
\begin{equation} \label{abcd}
    T^{QTM} (\la) = \begin{pmatrix}
                       A(\la) & B(\la) \\
                       C(\la) & D(\la)
                    \end{pmatrix}
\end{equation}
in `auxiliary space'.

\begin{example}
$X^{(1)} = \s^-$, $X^{(k - j + 1)} = \s^+$, and all other
$X^{(n)}$ are equal to the $2 \times 2$ unit matrix $I_2$. Then
(\ref{corloc}) turns into
\begin{equation} \label{spsmcor}
     \bigl\< \s_j^- \s_k^+ \bigr\>_T =
        \lim_{N \rightarrow \infty} \La_0^{j - k - 1} (0) \,
	\<\Ps_0| B(0) (A(0) + D(0))^{k - j - 1} C(0) |\Ps_0\> \epp
\end{equation}
\end{example}

\begin{example}
$X^{(1)} = X^{(2)} = \dots = X^{(k - j + 1)} = \re^{\ph e_2^2} =
\bigl(\begin{smallmatrix} 1 & 0 \\ 0 & \re^\ph \end{smallmatrix}\bigr)$
and $j = 1$, $k = m$. In this case (\ref{corloc}) turns into
\begin{equation} \label{genfun}
     \bigl\< \exp\bigl\{ \ph \tst{\sum_{n=1}^m} {e_n}_2^2 \bigr\}
        \bigr\>_T =
        \lim_{N \rightarrow \infty} \La_0^{- m} (0) \,
	\<\Ps_0| (A(0) + \re^\ph D(0))^m |\Ps_0\> \epp
\end{equation}
This special correlation function \cite{IzKo84} is a generating
function of the $\s^z$-$\s^z$ correlation functions which can be
expressed as
\begin{equation} \label{genfunappl}
     \bigl\< \s_1^z \s_m^z \bigr\>_T =
        (2 D_m^2 \6_\ph^2 - 4 D_m \6_\ph + 1) 
        \bigl\< \exp\bigl\{ \ph \tst{\sum_{n=1}^m} {e_n}_2^2 \bigr\}
                \bigr\>_T \Bigr|_{\ph = 0} \epc
\end{equation}
where $D_m$ is the `lattice derivative' defined on any complex sequence
$(a_n)_{n \in {\mathbb N}}$ by $D_m a_m = a_m - a_{m-1}$. The function
(\ref{genfun}) generates a number of other interesting correlation
functions that will be discussed below after we have derived an
integral representation for (\ref{genfun}).
\end{example}

The main goal of this work is to represent the generating function
(\ref{genfun}) as a multiple integral. The reason why we begin our
study of finite temperature correlation functions with the generating
function (\ref{genfun}) and not with any two-point function like
(\ref{spsmcor}) is that (\ref{genfun}) is slightly more simple than
the other interesting correlation functions, because the individual
factors in the product of operators on the right hand side of
(\ref{genfun}) are closely related to the quantum transfer matrix.

The expectation value on the right hand side of (\ref{genfun}) can
not easily be calculated directly. Instead, we shall introduce an
inhomogeneous version of it and shall perform the homogeneous limit
only at the end of our calculation. We introduce a `twisted quantum
transfer matrix' of the XXZ chain as
\begin{equation}
    t_\ph (\la) = A(\la) + \re^\ph D(\la) \epp
\end{equation}
For brevity we have suppressed the label QTM here. It follows from the
Yang-Baxter algebra (\ref{ybaop}) that the twisted quantum transfer
matrices $t_\ph (\la)$ for fixed $\ph$ form a commutative family, 
\begin{equation} \label{tphicom}
     [t_\ph (\la),t_\ph (\m)] = 0 \epp
\end{equation}
For any set $\{\x\} = \{\x_j\}_{j=1}^m$, $\x_j \in {\mathbb C}$, we
define a function
\begin{equation} \label{approxn}
     \PH_N (\ph|\{\x\})
        = \<\Ps_0|\bigl[ \prod_{j=1}^m t_\ph (\x_j) \bigr]
                  \bigl[ \prod_{j=1}^m t_0^{- 1} (\x_j) \bigr]|\Ps_0\>
\end{equation}
depending symmetrically (because of (\ref{tphicom})) on the $\x_j$.
Then the generating function (\ref{genfun}) can be calculated as
\begin{equation} \label{genfunaslim}
     \bigl\< \exp\bigl\{ \ph \tst{\sum_{n=1}^m} {e_n}_2^2 \bigr\}
        \bigr\>_T = \lim_{N \rightarrow \infty} \;
	            \lim_{\x_1, \dots, \x_m \rightarrow 0}
		    \PH_N (\ph|\{\x\}) \epp
\end{equation}
Our strategy below will be to first calculate $\PH_N (\ph|\{\x\})$
by means of the algebraic Bethe ansatz and then transform the resulting
expression into a form that will allow us to perform the homogeneous
limit and the Trotter limit on the right hand side of
(\ref{genfunaslim}).

\subsection{Algebraic Bethe ansatz for the quantum transfer matrix}
\label{dec:aba}
We shall now recall how to calculate spectrum and eigenvectors of
the quantum transfer matrix by means of the algebraic Bethe ansatz.
The result of an algebraic Bethe ansatz calculation is usually not
explicit but expresses eigenvalues and eigenvectors in terms of the
solutions of a set of Bethe ansatz equations. To actually solve the
Bethe ansatz equations is another problem we deal with in the next
section.

For the algebraic Bethe ansatz the form (\ref{rtt}) of the Yang-Baxter
algebra is more convenient than (\ref{ybaop}). Hence, we shall write
the monodromy matrix (\ref{monoqtm}) of the quantum transfer matrix
as a $2 \times 2$ matrix. For this purpose we first of all introduce
two $L$-matrices $L$ and $\tilde L$ setting
\begin{equation}
     {e_a}_\a^\be \, {L_j}^\a_\be (\la,\m) = R_{aj} (\la,\m) \epc \qd
        {e_a}_\a^\be \, \tilde{L_j}^\a_\be (- \m,\la)
          = R_{ja}^{t_1} (- \m,\la) \epp
\end{equation}
Comparing these equations with the explicit expression (\ref{rxxz})
for the $R$-matrix of the XXZ chain we obtain the explicit forms
of the two $L$-matrices,
\begin{subequations}
\label{lxxz}
\begin{align}
     L_j (\la,\m) & = \begin{pmatrix}
                       {e_j}_1^1 + b(\la,\m) \, {e_j}_2^2 &
		       c(\la,\m) \, {e_j}_2^1 \\
		       c(\la,\m) \, {e_j}_1^2 &
		       b(\la,\m) \, {e_j}_1^1 + {e_j}_2^2
                      \end{pmatrix} \epc \\[1ex]
     \tilde L_j (- \m,\la) & = \begin{pmatrix}
                              {e_j}_1^1 + b(- \m,\la) \, {e_j}_2^2 &
		              c(- \m,\la) \, {e_j}_1^2 \\
			      c(- \m,\la) \, {e_j}_2^1 &
			      b(- \m,\la) \, {e_j}_1^1 + {e_j}_2^2
                             \end{pmatrix} \epc
\end{align}
\end{subequations}
where we have employed the shorthand notations
\begin{equation}
     b(\la,\m) = \frac{\sh(\la - \m)}{\sh(\la - \m + \h)} \epc \qd
     c(\la,\m) = \frac{\sh(\h)}{\sh(\la - \m + \h)} \epp
\end{equation}
When expressed in terms of the $L$-matrices (\ref{lxxz}) the monodromy
matrix of the quantum transfer matrix becomes
\begin{equation}
     T^{QTM} (\la) = L_N \bigl( \la, \tst{\frac{\be}{N}} \bigr)
        \tilde L_{N-1} \bigl( - \tst{\frac{\be}{N}}, \la \bigr) \dots
        \tilde L_{1} \bigl( - \tst{\frac{\be}{N}}, \la \bigr) \epp
\end{equation}
This is now a $2 \times 2$ matrix of the form (\ref{abcd}) in `auxiliary
space' which satisfies the defining relations of the Yang-Baxter
algebra of the form (\ref{rtt}). Moreover, $T^{QTM} (\la)$ acts as
an upper triangular matrix on the vector
\begin{equation}
     |0\> = (e_1 \otimes e_2)^{\otimes \frac{N}{2}} =
        \underbrace{\tst{\binom{1}{0}} \otimes \tst{\binom{0}{1}}
	            \otimes \tst{\binom{1}{0}} \otimes \dots
		    \tst{\binom{0}{1}}}_{N \: {\rm factors}}
\end{equation}
which is obvious from the form of the $L$-matrices (\ref{lxxz}). More
precisely,
\begin{equation} \label{diagact}
     C(\la) |0\> = 0 \epc \qd
     A(\la) |0\> = a(\la) |0\> \epc \qd
     D(\la) |0\> = d(\la) |0\> \epp
\end{equation}
where
\begin{subequations}
\label{detpara}
\begin{align}
     a(\la) & = b \bigl( - \tst{\frac{\be}{N}},\la \bigr)^\frac{N}{2}
              = \biggl( \frac{\sh(\la + \frac{\be}{N})}
	                     {\sh(\la + \frac{\be}{N} - \h)}
		\biggr)^{\mspace{-6mu} \frac{N}{2}} \epc \\
     d(\la) & = b \bigl( \la,\tst{\frac{\be}{N}} \bigr)^\frac{N}{2}
              = \biggl( \frac{\sh(\la - \frac{\be}{N})}
	                     {\sh(\la - \frac{\be}{N} + \h)}
		\biggr)^{\mspace{-6mu} \frac{N}{2}} \epp
\end{align}
\end{subequations}
The first equation (\ref{diagact}) is sufficient for the algebraic
Bethe ansatz to work. It can be performed for arbitrary pseudo
vacuum eigenvalues $a(\la)$ and $d(\la)$ of the diagonal elements of
the monodromy matrix. These eigenvalues are often called `the
parameters of the algebraic Bethe ansatz'. Here they are given by
(\ref{detpara}). The parameters of the Bethe ansatz completely specify
the solution which for the general case can, for instance, be read up
in \cite{KBIBo}:

The vector
\begin{equation} \label{abaevec}
     |\{\la\}\> = |\{\la_j\}_{j=1}^M\> = B(\la_1) \dots B(\la_M) |0\>
\end{equation}
is an eigenvector of the quantum transfer matrix $t_0 (\la)$ if
the rapidities (or Bethe roots) $\la_j$, $j = 1, \dots, M$, satisfy
the system
\begin{equation} \label{bae}
     \frac{a(\la_j)}{d(\la_j)}
        = \prod_{\substack{k = 1 \\k \ne j}}^M
	  \frac{\sh(\la_j - \la_k + \h)}{\sh(\la_j - \la_k - \h)}
\end{equation}
of Bethe ansatz equations. The corresponding eigenvalue is
\begin{equation} \label{abaeval}
     \La(\la) = a(\la) \prod_{j=1}^M \frac{\sh(\la - \la_j - \h)}
                                          {\sh(\la - \la_j)}
	      + d(\la) \prod_{j=1}^M \frac{\sh(\la - \la_j + \h)}
		                          {\sh(\la - \la_j)} \epp
\end{equation}
Note that the `creation operators' $B(\la_j)$ mutually commute.
Therefore the eigenvector (\ref{abaevec}) is symmetric in the
rapidities, whence our notation $|\{\la\}\>$.

\subsection{Leading eigenvalue and auxiliary function}
\label{sec:auxfun}
In order to calculate the approximant $\PH_N (\ph|\{\x\})$ to the
generating function (\ref{genfun}) of the $\s^z$-$\s^z$ correlation
functions we need to know the specific solution $\{\la\}_{j=1}^M$ of
the Bethe ansatz equations (\ref{bae}) that determines the leading
eigenvalue $\La_0 (\la)$ and the corresponding eigenvector $|\Ps_0\>$.
This solution cannot be obtained in a closed analytic form. Even worse,
unlike in the case of the ordinary transfer matrix, the distribution
of the Bethe roots cannot be approximated by a smooth `density
function' in the Trotter limit, since they accumulate in the vicinity
of zero as $N$ increases. A way out of these difficulties was
proposed in \cite{Kluemper92,Kluemper93}: One associates an auxiliary
function
\begin{equation} \label{defa}
     \fa (\la) = \frac{d(\la)}{a(\la)} \prod_{k = 1}^{N/2}
	         \frac{\sh(\la - \la_k + \h)}{\sh(\la - \la_k - \h)}
\end{equation}
with the solution $\{\la\} = \{\la_k\}_{k=1}^\frac{N}{2}$ of the Bethe
ansatz equations that determines the leading eigenvalue of the quantum
transfer matrix. It then turns out that the function $\fa (\la)$ is
sufficiently well determined by the gross properties of $\{\la\}$.
Knowing that the leading eigenvalue for fixed Trotter number $N$ is
determined by
\begin{enumerate}
\item
$M = N/2$ Bethe roots\footnote{This was already used in the definition
(\ref{defa}) of $\fa (\la)$. For technical reasons we assume that
$N/4 \in {\mathbb N}$.} which
\item
are mutually distinct and
\item
lie all on the imaginary axis
\end{enumerate}
we can derive a non-linear integral equation which (alternatively)
defines $\fa (\la)$. The details of the derivation have been described
on several earlier occasions \cite{Kluemper04}. Here we only note that
(i)-(iii) can be established `perturbatively' in the high temperature
limit and have been verified numerically for a wide range of Trotter
numbers and temperatures. Conditions (i)-(iii) determine the analytical
properties of the auxiliary function $\fa (\la)$.
\begin{enumerate}
\item
The auxiliary function $\fa$ is meromorphic and periodic with period
$\i \p$ in imaginary direction.
\item
It has $3N/2$  poles (including multiplicities) in the fundamental
domain $D = \{z \in \nolinebreak {\mathbb C}| - \p/2 < |\Im z| \le
\p/2\}$: $N/2$ simple poles at $\la_j + \h$, $j = 1, \dots, N/2$;
an $N/2$-fold pole at $\la = - \be/N$; and an $N/2$-fold pole at
$\la = \be/N - \h$.
\item
The meromorphic function $1 + \fa$ has $3N/2$ zeros in $D$: $N/2$
simple zeros at the Bethe ansatz roots,
\begin{equation} \label{bafroma}
     1 + \fa (\la_j) = 0 \epc
\end{equation}
clustering close to $\la = 0$ as $\be \rightarrow 0$; $N/2$ zeros
which are close to $\h$ as $\be \rightarrow 0$; and $N/2$ zeros which
accumulate at $- \h$ for $\be \rightarrow 0$. We assume of the latter
two kinds of zeros that the modulus of their real part stays
larger than $\h/2$ for all finite temperatures.
\end{enumerate}

\begin{figure}

\begin{center}

\epsfxsize 9cm
\epsffile{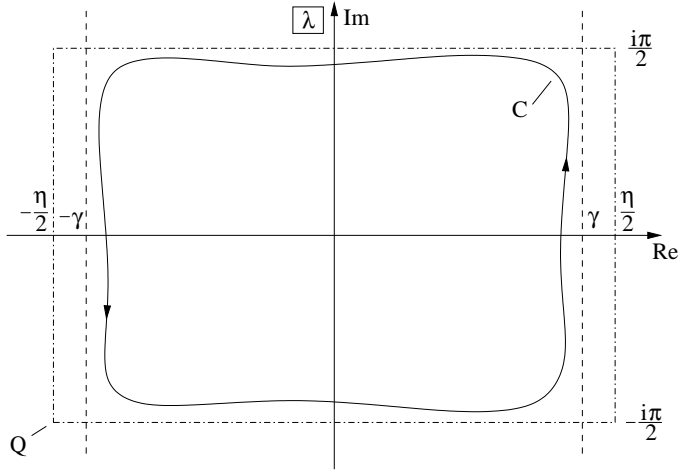}

\caption{\label{fig:cancon} The canonical contour ${\cal C}$ for the
off-critical regime $\D > 1$.}
\end{center}

\end{figure}

Once the analytical properties of $\fa$ and $1 + \fa$ are established
it is not difficult to derive the following integral equation for
the auxiliary function,
\begin{equation} \label{nlien}
     \ln \fa (\la) = \ln \biggl[
                      \frac{\sh(\la - \frac{\be}{N})
		            \sh(\la + \frac{\be}{N} + \h)}
			   {\sh(\la + \frac{\be}{N})
			    \sh(\la - \frac{\be}{N} + \h)}
			    \biggr]^{\mspace{-3mu} \frac{N}{2}}
	\mspace{-3mu}
        - \int_{\cal C} \frac{d \om}{2 \p \i} \,
	  \frac{\sh (2 \h) \ln (1 + \fa (\om))}
	       {\sh(\la - \om + \h) \sh(\la - \om - \h)},
\end{equation}
where (for $\D > 1$) the contour ${\cal C}$ (see figure
\ref{fig:cancon}) is a rectangular contour with edges parallel to the
real axis at $\pm \i  \p/2$ and to the imaginary axis at $\pm \g$ where
$0 < \g < \h/2$. Equation (\ref{nlien}) is valid inside the rectangle
$Q$ defined by $|\Re \la| < \h/2$ and $|\Im \la| < \p/2$. This means
that it defines the auxiliary function $\fa (\la)$ for all other $\la$
in $Q$ once it has been calculated on~${\cal C}$.
In particular, one may calculate the zeros of $1 + \fa$ inside
${\cal C}$ which, by construction (see (\ref{bafroma})), are those
solutions of the Bethe ansatz equations that determine the leading
eigenvalue and the corresponding eigenvector. We will, however, not
use equation (\ref{nlien}) in this way. It turns out that all 
quantities we are interested in, the free energy and the generating
function of the $\s^z$-$\s^z$ correlation functions in particular,
can be expressed as integrals over the auxiliary function $\fa$ on
the contour~${\cal C}$.

It is very important that the contour ${\cal C}$ of the integral on
the right hand side of (\ref{nlien}) does not depend on $N$. The
Trotter number $N$ merely appears as a parameter in the inhomogeneity
of the integral equation. Hence, the Trotter limit can be performed in
(\ref{nlien}). We obtain
\begin{equation} \label{nlie}
     \ln \fa (\la) = - \frac{2J \sh^2 (\h)}{T \sh(\la) \sh(\la + \h)}
                     - \int_{\cal C} \frac{d \om}{2 \p \i} \,
	               \frac{\sh (2 \h) \ln (1 + \fa (\om))}
		            {\sh(\la - \om + \h)
			            \sh(\la - \om - \h)} \epp
\end{equation}
Here we already took into account the rescaling of the Hamiltonian (see
(\ref{rescaleham}) and (\ref{stform})) and introduced the temperature
$T$ as
\begin{equation}
     T = \frac{2J \sh(\h)}{\be} \epp
\end{equation}

The auxiliary function $\fa$ encodes the information about the
location of those Bethe roots that determine the leading eigenvalue
of the quantum transfer matrix and the corresponding eigenvector.
Using the results of sections \ref{sec:qtm} and \ref{sec:corqtm} we
conclude that $\fa$ thus encodes not only the complete information 
about the free energy but also all the `thermodynamically relevant
information' about the correlation functions of local operators.

\subsection{Including the magnetic field}
\label{sec:magfie}
Before coming to the calculation of physical quantities in the
next sections we have to discuss the issue of how to include a
(longitudinal) magnetic field into the formalism. For this purposes
we have to replace the statistical operator $\re^{- \be H}$ by
$\re^{- (\be H - h S^z/T)}$, where $h$ denotes the magnetic field and
\begin{equation}
     S^z = \tst{\2} \sum_{j=1}^L \s_j^z
\end{equation}
is the operator of the $z$-component of the total spin.

The reason why this is a rather simple task lies in the commutativity
of the $R$-matrix (\ref{rxxz}) with the tensorial square of the
diagonal matrix $\Th = \diag(\re^\th, \re^{- \th})$,
\begin{equation} \label{rcomthetaint}
     [R(\la,\m),\Th \otimes \Th] = 0 \epp
\end{equation}
Taking the derivative of this equation with respect to $\th$ at
$\th = 0$ we obtain
\begin{equation}
     [R(\la,\m),\Th \otimes I_2 + I_2 \otimes \Th] = 0 \epp
\end{equation}
The latter equation implies that the ordinary transfer matrix and
hence the Hamiltonian commutes with $S^z$. On the other hand
(\ref{rcomthetaint}) is equivalent to
\begin{equation} \label{threpyba}
     \check R (\la,\m) (\Th \otimes \Th) =
        (\Th \otimes \Th) \check R (\la,\m) \epc
\end{equation}
which means that $\Th$ is a spectral parameter independent
representation of the Yang-Baxter algebra with $R$-matrix (\ref{rxxz}).
It follows that
\begin{equation}
     \lim_{N \rightarrow \infty} \r_{N,L} \re^\frac{h S^z}{T}
        = \re^{- (\be H - h S^z/T)} \epc
\end{equation}
while, on the other hand,
\begin{equation}
     \r_{N,L} \, e^\frac{h S^z}{T} = \tr_{\bar 1, \dots, \bar N} \;
        T_1^{QTM} (0) \Bigl( \begin{smallmatrix} \re^{h/2T} & 0 \\
	                     0 & \re^{- h/2T} \end{smallmatrix} \Bigr)_1
			     \dots
        T_L^{QTM} (0) \Bigl( \begin{smallmatrix} \re^{h/2T} & 0 \\
	                     0 & \re^{- h/2T} \end{smallmatrix} \Bigr)_L
			     \epc
\end{equation}
where $T_j^{QTM} (\la)$ is the monodromy matrix corresponding to the
quantum transfer matrix as defined for zero magnetic field.

Thus, we can include a longitudinal magnetic field by simply replacing
\begin{equation}
     T^{QTM} (\la) \rightarrow
        T^{QTM} (\la) \Bigl( \begin{smallmatrix} \re^{h/2T} & 0 \\
	                     0 & \re^{- h/2T} \end{smallmatrix} \Bigr)
\end{equation}
in our former calculations. The important point here is that, because
of (\ref{threpyba}), the redefined monodromy matrix is still a
representation of the Yang-Baxter algebra with the same $R$-matrix
(\ref{rxxz}). Therefore our former calculations go through without
almost any modification. We simply replace $a(\la)$ by $a(\la)
\re^{h/2T}$ and $d(\la)$ by $d(\la) \re^{-h/2T}$ in (\ref{bae})-%
(\ref{defa}). The only point one then has to take care of is that the
loci of the Bethe roots that describe the leading eigenvalue are now
off the imaginary axis. As a consequence the parameter $\g$ appearing
in our definition of the path above has to be chosen large enough (but
still between zero and $\h/2$). Then the only modification of the
integral equations (\ref{nlien}) and (\ref{nlie}) is the appearance of
an extra term $- h/T$ on the right hand side. For example,
(\ref{nlie}) turns into
\begin{equation} \label{nlieh}
     \ln \fa (\la) = - \frac{h}{T}
                     - \frac{2J \sh^2 (\h)}{T \sh(\la) \sh(\la + \h)}
                     - \int_{\cal C} \frac{d \om}{2 \p \i} \,
	               \frac{\sh (2 \h) \ln (1 + \fa (\om))}
		            {\sh(\la - \om + \h)
			            \sh(\la - \om - \h)} \epp
\end{equation}
This is the non-linear integral equation for the auxiliary function
$\fa$ of the XXZ chain exposed to a finite, longitudinal magnetic
field $h$.

It is often useful (in particular when it comes to numerical
calculations) to consider the auxiliary function $\fa$ together with
its `dual' $\faq = 1/\fa$. As we can see from equation (\ref{bafroma})
the function $\faq$ is equally useful in determining the Bethe roots,
for $1 + \fa (\la) = 0$ is equivalent to $1 + \faq (\la) = 0$. An
integral equation for $\faq$ can be directly obtained from
(\ref{nlieh}). For this purpose we choose $\la$ and $\om$ from inside
the rectangle $Q$. Then $|\Re \la|, |\Re \om| < \h/2$ and hence
$\la - \om \pm \h \ne 0$. Taking this into account we change the
free variable in equation (\ref{nlieh}) from $\la$ to $\om$, multiply
it by $\sh(2\h)/2 \p \i \, \sh(\la - \om + \nolinebreak \h)
\sh(\la - \om - \h)$ and integrate over ${\cal C}$. The resulting
formula is
\begin{equation}
     \int_{\cal C} \frac{d \om}{2 \p \i} \,
        \frac{\sh (2 \h) \ln (\fa (\om))}
	     {\sh(\la - \om + \h) \sh(\la - \om - \h)}
     = - \frac{2J \sh (\h) \sh(2\h)}{T \sh(\la + \h) \sh(\la - \h)} \epp
\end{equation}
We use it in (\ref{nlieh}) and obtain an integral equation for $\faq$:
\begin{equation} \label{nliehaq}
     \ln \faq (\la) = \frac{h}{T}
                      - \frac{2J \sh^2 (\h)}{T \sh(\la) \sh(\la - \h)}
                      + \int_{\cal C} \frac{d \om}{2 \p \i} \,
	                \frac{\sh (2 \h) \ln (1 + \faq (\om))}
			     {\sh(\la - \om + \h)\sh(\la - \om - \h)}
			      \epp
\end{equation}
Comparing this with (\ref{nlieh}) we see that the formal difference
consists in replacing $h$ and $\h$ by $- h$ and $- \h$.

\subsection{Free energy and magnetization}
\label{sec:freemandela}
Starting from the expression (\ref{abaeval}) (properly modified such
as to include the magnetic field) we can derive expressions that
represent the free energy per lattice site $f$ as integrals either
over $\fa$ or $\faq$ (for details see \cite{Kluemper04}),
\begin{equation} \label{freee}
     f (h,T) = - \frac{h}{2} - T \mspace{-5mu} \int_{\cal C} \frac{d \om}{2 \p \i} \,
             \frac{\sh (\h) \ln (1 + \fa (\om))}
	          {\sh(\om) \sh(\om + \h)}
       = \frac{h}{2} + T \mspace{-5mu} \int_{\cal C} \frac{d \om}{2 \p \i} \,
             \frac{\sh (\h) \ln (1 + \faq (\om))}
	          {\sh(\om) \sh(\om - \h)}.
\end{equation}

From these equations the thermodynamic properties of the XXZ chain
can be calculated. For example, the magnetization per unit length of
the chain at finite magnetic field $h$ and finite temperature $T$,
defined as
\begin{equation}
     m(h,T) = \lim_{L \rightarrow \infty} \frac{\<S^z\>_{h,T}}{L}
\end{equation}
can be calculated by means of the thermodynamic relation
\begin{equation}
     m(h,T) = - \, \frac{\6 f(h,T)}{\6 h} \epp
\end{equation}
Introducing the function
\begin{equation} \label{defmagdens}
     \s (\la) = - \, \frac{T \6_h \fa (\la)}{\fa (\la)}
              = \frac{T \6_h \faq (\la)}{\faq (\la)}
\end{equation}
and taking the derivative of (\ref{freee}) with respect to the
magnetic field we obtain
\begin{equation} \label{msdens}
     m(h,T) = - \2 - \int_{\cal C} \frac{d \om}{2 \p \i} \,
             \frac{\sh (\h)}{\sh(\om) \sh(\om - \h)}
             \frac{\s (\om)}{1 + \fa(\om)} \epp
\end{equation}
Instead of calculating the explicit $h$-dependence of the functions
$\fa$ or $\faq$ and from this the function $\s$, we may calculate it
from the linear integral equation
\begin{equation} \label{intsigma}
     \s (\la) = 1 + \int_{\cal C} \frac{d \om}{2 \p \i} \,
	            \frac{\sh (2 \h)}
		         {\sh(\la - \om + \h)\sh(\la - \om - \h)}
	            \frac{\s (\om)}{1 + \fa (\om)}
\end{equation}
obtained from (\ref{nliehaq}) by taking the derivative with respect
to $h$. Equation (\ref{intsigma}) is numerically better conditioned
than (\ref{defmagdens}) and therefore particularly useful if one is
interested in numbers.

For later use we derive another formula for the magnetization using an
analogy with the dressed function formalism for the density functions
at zero temperature. We define a function $G(\la)$ as the solution
of the linear integral equation
\begin{equation} \label{defghom}
     G (\la) = \frac{\sh (\h)}{\sh(\la) \sh(\la - \h)}
                  + \int_{\cal C} \frac{d \om}{2 \p \i} \,
	            \frac{\sh (2 \h)}
		         {\sh(\la - \om + \h)\sh(\la - \om - \h)}
	            \frac{G (\om)}{1 + \fa (\om)} \epp
\end{equation}
Then it is not difficult to show that
\begin{equation}
     \int_{\cal C} \frac{d \om}{2 \p \i} \,
         \frac{\sh (\h)}{\sh(\om) \sh(\om - \h)}
         \frac{\s (\om)}{1 + \fa(\om)} =
     \int_{\cal C} \frac{d \om}{2 \p \i} \,
         \frac{G (\om)}{1 + \fa(\om)} \epc
\end{equation}
and it follows from (\ref{msdens}), (\ref{defghom}) that
\begin{equation} \label{mgdens}
     m(h,T) = - \2 - \int_{\cal C} \frac{d \om}{2 \p \i} \,
             \frac{G (\om)}{1 + \fa(\om)}
            = \2 + \int_{\cal C} \frac{d \om}{2 \p \i} \,
             \frac{G (\om)}{1 + \faq (\om)} \epp
\end{equation}

\section{Results} \label{sec:result}
\subsection{Combinatorics}
We shall now derive our main result which is an integral formula for
the generating function (\ref{genfun}) of the $\s^z$-$\s^z$ correlation
functions of the XXZ chain. The general strategy we are going to
pursue is to first calculate the inhomogeneous, finite Trotter
number approximant $\PH_N (\ph|\{\x\})$, equation (\ref{approxn}),
and then transform it into a form such that the limits in 
(\ref{genfunaslim}) can be performed.

The Bethe ansatz eigenvectors (\ref{abaevec}) are not `normalized'.
Hence, $\PH_N (\ph|\{\x\})$, when expressed through the Bethe ansatz
eigenvectors takes the form
\begin{equation} \label{phnaba}
     \PH_N (\ph|\{\x\}) = \frac{\bigl\< \{\la\} \bigr|
                          \bigl[\prod_{j=1}^m t_\ph(\x_j)\bigr]
			  \bigl[\prod_{j=1}^m t_0^{-1}(\x_j)\bigr]
			  \bigl| \{\la\} \bigr\>}
			       {\bigl\< \{\la\} \big| \{\la\} \bigr\>}
			       \epc
\end{equation}
where the $\la_j$ in $\{\la\}$ are those that characterize the leading
eigenvalue. The dual $\<\{\la\}|$ to $|\{\la\}\>$ can be expressed
through the action of the operators $C(\la)$ (see (\ref{abcd})) on the
dual pseudo vacuum $\<0|$ which is the transposed of $|0\>$,
\begin{equation}
     \<\{\la\}| = \<0| C(\la_1) \dots C(\la_{N/2}) \epp
\end{equation}

In order to evaluate the numerator in (\ref{phnaba}) we calculate
in the first instance the left action of $\prod_{j=1}^m t_\ph(\x_j)$
on $\<\{\la\}|$. In this task we can largely follow \cite{KMST02a}.
The basic idea is to use the commutation relations
\begin{subequations}
\label{cadcom}
\begin{align}
     C(\m) A(\la) & = f(\m,\la) A(\la) C(\m) - g(\m,\la) A(\m) C(\la)
                      \epc \\[.5ex]
     C(\m) D(\la) & = f(\la,\m) D(\la) C(\m) - g(\la,\m) D(\m) C(\la)
\end{align}
\end{subequations}
contained in the defining relations (\ref{rtt}) of the Yang-Baxter
algebra in order to move the operators $A$ and $D$ (implicit in the
expression $t_\ph$) in
\begin{equation} \label{todo}
     \<0| C(\la_1) \dots C(\la_{N/2}) t_\ph (\x_1) \dots t_\ph (\x_m)
\end{equation}
to the left and replace them with their pseudo vacuum expectation values
$a$ and $d$ (which are the same for right action on the pseudo vacuum
and left action on its dual (see (\ref{lxxz}))). Our notation employed
in (\ref{cadcom}) is
\begin{equation}
     f(\la,\m) = \frac{\sh(\la - \m + \h)}{\sh(\la - \m)} \epc \qd
     g(\la,\m) = \frac{\sh(\h)}{\sh(\la - \m)} \epp
\end{equation}
Except for the commutation relations (\ref{cadcom}) one needs in the
actual calculation the mutual commutativity of the families of
operators $C(\la)$ and $t_\ph (\la)$, respectively,
\begin{equation} \label{ctcom}
     [C(\la),C(\m)] = [t_\ph (\la), t_\ph (\m)] = 0 \epc
\end{equation}
which also follows from (\ref{rtt}).

When we apply (\ref{cadcom}) successively in (\ref{todo}) in order to
move the operators $A$ and $D$ to the very left, then the number of
operators $C$ per term in a single application of (\ref{cadcom}) is
conserved. Merely the arguments of the commuted operators are exchanged
in every possible way. To state this more formally we need to introduce
several set theoretical notations.

We shall deal with finite sets of complex numbers such as
$\{\la_j\}_{j=1}^M$. For such sets we already introduced the short
hand notation $\{\la\} = \{\la_j\}_{j=1}^M$. Similarly we shall
write$\{\x\} = \{\x_j\}_{j=1}^M$ etc. The number of elements in these
sets will be somewhat loosely denoted as $|\la|$, $|\x|$, \dots We
further need partitions of sets into disjoint subsets. By $p_2 \{\la\}$
we shall denote the set of all ordered pairs $(\{\la^+\},\{\la^-\})$ of
subsets $\{\la^+\}, \{\la^-\} \subset \{\la\}$ which satisfy $\{\la^+\}
\cup \{\la^-\} = \{\la\}$ and $\{\la^+\} \cap \{\la^-\} = \emptyset$.
The elements of the subsets $\{\la^\pm\}$ will be denoted $\la_j^\pm$,
such that we can write $\{\la^\pm\} = \{\la_j^\pm\}_{j=1}^{|\la^\pm|}$
if we want to emphasize the number of elements in the subsets. Using
our new notation we conclude with (\ref{cadcom}) and (\ref{ctcom}) that
\begin{equation} \label{twistact}
     \<\{\la\}| \prod_{j=1}^{|\x|} t_\ph (\x_j)
        = \mspace{-18mu}
	  \sum_{\substack{(\{\la^+\},\{\la^-\}) \in p_2 \{\la\}\\
           (\{\x^+\},\{\x^-\}) \in p_2 \{\x\}\\
	   |\x^+| + |\la^-| = N/2}} \:
	   R(\{\x^+\}|\{\x^-\}|\{\la^+\}|\{\la^-\}) \,
	   \<\{\x^+\} \cup \{\la^-\}|
\end{equation}
with numerical coefficients $R$ depending on the sets $\{\x^\pm\}$,
$\{\la^\pm\}$. These coefficients can be calculated in the general
situation of arbitrary vacuum expectation values $a$, $d$ of the
diagonal elements of the monodromy matrix and for arbitrary sets
of mutually distinct complex numbers $\{\x\}$, $\{\la\}$, where the
$\la_j$ do not necessarily satisfy the Bethe ansatz equations.
\begin{lemma} \label{lem:multactt}
\cite{KMST02a} The coefficients $R(\{\x^+\}|\{\x^-\}|\{\la^+\}|
\{\la^-\})$ in (\ref{twistact}) are given by the formula
\begin{multline}
     R(\{\x^+\}|\{\x^-\}|\{\la^+\}|\{\la^-\}) \\=
        S(\{\x^+\}|\{\la^+\}|\{\la^-\})
	\prod_{j=1}^{|\x^-|} \biggl\{ a(\x_j^-)
	 \Bigl[ \prod_{k=1}^{|\x^+|} f(\x_k^+,\x_j^-) \Bigr]
	 \Bigl[ \prod_{l=1}^{|\la^-|} f(\la_l^-,\x_j^-) \Bigr] \\
	 + \re^\ph d(\x_j^-)
	 \Bigl[ \prod_{k=1}^{|\x^+|} f(\x_j^-,\x_k^+) \Bigr]
	 \Bigl[ \prod_{l=1}^{|\la^-|} f(\x_j^-,\la_l^-) \Bigr]
	 \biggr\} \epc
\end{multline}
where the so-called highest coefficient $S(\{\x^+\}|\{\la^+\}|
\{\la^-\})$ can be expressed as the ratio of two determinants:
\begin{equation}
     S(\{\x^+\}|\{\la^+\}|\{\la^-\})
        = \frac{\det \widehat{M} (\la_j^+,\x_k^+)}
	       {\det V (\la_j^+,\x_k^+)}
\end{equation}
with
\begin{align}
     V (\la_j^+,\x_k^+) & = \frac{1}{\sh(\x_k^+ - \la_j^+)} \epc \\
     \widehat{M} (\la_j^+,\x_k^+)
        & = a(\la_j^+) t(\x_k^+,\la_j^+)
	    \Bigl[ \prod_{l=1}^{|\x^+|} f(\x_l^+,\la_j^+) \Bigr]
	    \Bigl[ \prod_{m=1}^{|\la^-|} f(\la_m^-,\la_j^+) \Bigr]
	    \notag \\ & \mspace{99mu}
	    - \re^\ph d(\la_j^+) t(\la_j^+,\x_k^+)
	    \Bigl[ \prod_{l=1}^{|\x^+|} f(\la_j^+,\x_l^+) \Bigr]
	    \Bigl[ \prod_{m=1}^{|\la^-|} f(\la_j^+,\la_m^-) \Bigr]
	    \epc \\
     t (\la, \x) & = \frac{\sh(\h)}{\sh(\la - \x)\sh(\la - \x + \h)}
                     \epp
\end{align}
\end{lemma}

We may now insert (\ref{twistact}) into (\ref{phnaba}) and use the
fact, that the vector $|\{\la\}\>$ in (\ref{phnaba}) is the
eigenvector of the quantum transfer matrix that belongs to the
leading eigenvalue
\begin{equation} \label{leval}
     \La_0 (\la) = a(\la) \prod_{j=1}^{N/2} f(\la_j,\la)
                 + d(\la) \prod_{j=1}^{N/2} f(\la,\la_j)
\end{equation}
(see (\ref{abaeval})). We obtain
\begin{equation} \label{twistactcalc}
     \PH_N (\ph|\{\x\})
        = \mspace{-18mu}
	  \sum_{\substack{(\{\la^+\},\{\la^-\}) \in p_2 \{\la\}\\
           (\{\x^+\},\{\x^-\}) \in p_2 \{\x\}\\
	   |\x^+| + |\la^-| = N/2}}
	   \frac{R(\{\x^+\}|\{\x^-\}|\{\la^+\}|\{\la^-\}) \,
	         \<\{\x^+\} \cup \{\la^-\}|\{\la\}\>}
                {\Bigl[ \prod_{j=1}^{|\x|} \La_0 (\x_j) \Bigr]
		 \<\{\la\}|\{\la\}\>} \epc
\end{equation}
and only the ratio $\<\{\x^+\} \cup \{\la^-\}|\{\la\}\>/
\<\{\la\}|\{\la\}\>$ remains to be calculated. But here another
combinatorial result for the XXZ chain applies.
\begin{lemma} \label{lem:slavnov}
\cite{Slavnov89} Let $\{\la_j\}_{j=1}^M$ a solution of the Bethe
ansatz equations (\ref{bae}) and $\{\m_k\}_{k=1}^M$ any set of distinct
complex numbers. Then
\begin{multline} \label{slavnov}
     \<0|C(\m_1) \dots C(\m_M)B(\la_1) \dots B(\la_M)|0\>\\
        = \frac{\bigl[ \prod_{j=1}^M d(\la_j) a(\m_j) \bigr]
	        \prod_{j, k = 1}^M \sh(\la_j - \m_k + \h)}
               {\prod_{1 \le j < k \le M}
	        \sh(\la_j - \la_k)\sh(\m_k - \m_j)} \;
          \det \widehat{N} (\la_j,\m_k) \epc
\end{multline}
where
\begin{equation} \label{deftilden}
     \widehat{N} (\la_j,\m_k) = t(\la_j,\m_k) - t(\m_k,\la_j)
                                  \frac{d(\m_k)}{a(\m_k)}
				  \prod_{l=1}^M
				  \frac{f(\m_k,\la_l)}{f(\la_l,\m_k)}
				  \epp
\end{equation}
\end{lemma}

In order to apply lemma \ref{lem:slavnov} to our task of calculating
$\<\{\x^+\} \cup \{\la^-\}|\{\la\}\>/ \<\{\la\}|\{\la\}\>$ we first
insert the set of Bethe roots that characterize the leading eigenvalue
of the quantum transfer matrix into (\ref{slavnov}), (\ref{deftilden}).
Then $M = N/2$, and the right hand side of (\ref{deftilden}) can be
expressed in terms of the auxiliary function (\ref{defa}),
\begin{equation} \label{tildenaf}
     \widehat{N} (\la_j,\m_k)
        = t(\la_j,\m_k) - t(\m_k,\la_j) \fa(\m_k) \epp
\end{equation}
Taking account of the Bethe ansatz equations $\fa(\la_k) = -1$ it is
easy to study the limit $\m_k \rightarrow \la_k$ in (\ref{slavnov}):
The prefactor is regular in this limit. Concerning the matrix
$\widehat{N}$ only the $k$th column is affected,
\begin{equation} \label{normlim}
     \lim_{\m_k \rightarrow \la_k} \widehat{N} (\la_j,\m_k)
        = \de^j_k \; \frac{\fa' (\la_k)}{\fa (\la_k)} +
	  \frac{\sh(2\h)}
	       {\sh(\la_j - \la_k + \h)\sh(\la_j - \la_k - \h)} \epp
\end{equation}
Adopting for a while the notation
\begin{equation} \label{doptnot}
\begin{split}
     \tilde \la_j & = \begin{cases}
                      \la_j^+ & \text{for $j = 1, \dots, |\x^+|$}
		         \\[.5ex]
		      \la_{j - |\x^+|}^- & \text{for $j = |\x^+| + 1,
		         \dots, N/2,$}
                      \end{cases} \\[1ex]
     \tilde \x_j & = \x_j^+ \qd \text{for $j = 1, \dots, |\x^+| = n$,}
\end{split}
\end{equation}
we infer from (\ref{slavnov})-(\ref{normlim}) that
\begin{multline} \label{preratio}
     \frac{\<\{\x^+\} \cup \{\la^-\}|\{\la\}\>}
          {\<\{\la\}|\{\la\}\>} =
	  \biggl[ \prod_{j=1}^n \frac{a(\tilde \x_j)}{a(\tilde \la_j)}
	         \biggr]
	  \biggl[ \prod_{j = n + 1}^{N/2} \prod_{k=1}^n
	         \frac{f(\tilde \la_j, \tilde \x_k)}
	              {f(\tilde \la_j, \tilde \la_k)}
	         \biggr] \\
	  \biggl[ \prod_{j,k=1}^n
	         \frac{\sh(\tilde \la_j - \tilde \x_k + \h)}
	              {\sh(\tilde \la_j - \tilde \la_k + \h)}
	         \biggr]
	  \biggl[ \prod_{1 \le j < k \le n}
	         \frac{\sh(\tilde \la_j - \tilde \la_k)}
	              {\sh(\tilde \x_j - \tilde \x_k)}
	         \biggr]
          \frac{\det N_n}{\det N_0} \epc
\end{multline}
where
\begin{equation}
     {N_n}^j_k = \begin{cases}
        t(\tilde \la_j,\tilde \x_k)
	   - t(\tilde \x_k,\tilde \la_j) \fa(\tilde \x_k)
	   & k = 1, \dots, n \\[1.5ex]
        \dst{\de^j_k \; \frac{\fa' (\tilde \la_k)}{\fa (\tilde \la_k)}
	   + \frac{\sh(2\h)}
	          {\sh(\tilde \la_j - \tilde \la_k + \h)
		   \sh(\tilde \la_j - \tilde \la_k - \h)}}
           & k = n + 1, \dots, N/2 \epp
                 \end{cases}
\end{equation}
The observation that the columns of $N_n$ and $N_0$ are identical for
$k > n$ allows one to simplify the ratio of the two determinants on
the right hand side of (\ref{preratio}). In the appendix we prove the
following
\begin{lemma} \label{lem:detrat}
The ratio of the two determinants on the right hand side of
(\ref{preratio}) is proportional to the determinant of an $n \times n$
matrix,
\begin{equation}
          \frac{\det N_n}{\det N_0}
	     = \biggl[ \prod_{l=1}^n
	       \frac{1 + \fa(\x_l^+)}{\fa'(\la_l^+)} \biggr]
	       \det G(\la_j^+,\x_k^+) \epc
\end{equation}
where the function $G(\la,\x)$ is the solution of the linear integral
equation
\begin{equation} \label{defg}
     G(\la,\x) = t(\x,\la)
                  + \int_{\cal C} \frac{d \om}{2 \p \i} \,
	            \frac{\sh (2 \h)}
		         {\sh(\la - \om + \h)\sh(\la - \om - \h)}
	            \frac{G (\om,\x)}{1 + \fa (\om)}
\end{equation}
to be solved on the same canonical contour {\cal C} as the non-linear
integral equation for the auxiliary function $\fa$.
\end{lemma}
Notice that $G(\la,\x)$ is an inhomogeneous generalization of the
`finite temperature density function' $G(\la)$, equation
(\ref{defghom}), that determines the magnetization (\ref{mgdens}).

Using the lemma in (\ref{preratio}) and switching back to our original
notation we obtain the following result,
\begin{multline} \label{ratio}
     \frac{\<\{\x^+\} \cup \{\la^-\}|\{\la\}\>}{\<\{\la\}|\{\la\}\>}
        = \biggl[ \prod_{j=1}^{|\x^+|}
	          \frac{a(\x_j^+)(1 + \fa (\x_j^+))}
		       {a(\la_j^+) \fa' (\la_j^+)} \biggr]
	  \biggl[ \prod_{j = 1}^{|\la^-|} \prod_{k=1}^{|\x^+|}
	         \frac{f(\la_j^-, \x_k^+)}{f(\la_j^-, \la_k^+)} \biggr]
		 \\
	  \biggl[ \prod_{j,k=1}^{|\x^+|}
	         \frac{\sh(\la_j^+ - \x_k^+ + \h)}
	              {\sh(\la_j^+ - \la_k^+ + \h)} \biggr]
	  \biggl[ \prod_{1 \le j < k \le |\x^+|}
	         \frac{\sh(\la_j^+ - \la_k^+)}{\sh(\x_j^+ - \x_k^+)}
		 \biggr]
	  \det G(\la_j^+,\x_k^+) \epp
\end{multline}
Finally, we use lemma \ref{lem:multactt} and equations (\ref{leval}),
(\ref{ratio}) in (\ref{twistactcalc}) and remove the explicit
dependence on the $\la_j^-$ from the resulting expression. The latter
is possible by proper use of the auxiliary function $\fa$ and the Bethe
ansatz equations (\ref{bae}). The calculation is elementary but
slightly tedious. We obtain
\begin{lemma}
The finite Trotter number approximant $\PH_N (\ph|\{\x\})$, equation
(\ref{phnaba}), has the following representation as a sum over
partitions of the Bethe roots $\{\la\}$, that characterize the leading
eigenvalue, and of the inhomogeneity parameters $\{\x\}$,
\begin{equation} \label{sumrep}
     \PH_N (\ph|\{\x\})
        = \mspace{-18mu}
	  \sum_{\substack{(\{\la^+\},\{\la^-\}) \in p_2 \{\la\}\\
           (\{\x^+\},\{\x^-\}) \in p_2 \{\x\}\\
	   |\x^+| + |\la^-| = N/2}}
           \frac{Y_{|\x^+|} (\{\la^+\}|\{\x^+\}) \;
	         Z_{|\x^+|} (\{\la^+\}|\{\x^+\}|\{\x^-\})}
                {\Bigr[ \prod_{j=1}^{|\x^+|} \fa'(\la_j^+) \Bigr]
                 \Bigr[ \prod_{j=1}^{|\x^-|}
		        \bigl(1 + \fa(\x_j^-)\bigr) \Bigr]} \epc
\end{equation}
where the two functions $Y_n (\{\la^+\}|\{\x^+\})$ and $Z_n
(\{\la^+\}|\{\x^+\}|\{\x^-\})$ are defined as follows:
\begin{multline}
     Y_n (\{\la^+\}|\{\x^+\})
        = \biggl[ \prod_{j=1}^n \frac{\fbq (\la_j^+)}{\fbq' (\x_j^+)}
	          \biggr]
          \biggl[ \prod_{j,k=1}^n
	     \frac{\sh(\la_j^+ - \x_k^+ + \h)\sh(\la_j^+ - \x_k^+ - \h)}
	          {\sh(\x_j^+ - \x_k^+ + \h)\sh(\la_j^+ - \la_k^+ - \h)}
	          \biggr] \\
          \det \widetilde{M} (\la_j^+,\x_k^+) \; \det G(\la_j^+,\x_k^+)
\end{multline}
with
\begin{align}
     \fbq (\la) & = \prod_{k=1}^m \frac{1}{f(\la,\x_k)} \epc \\
     \widetilde{M} (\la_j^+,\x_k^+) & = t(\x_k^+,\la_j^+) +
        t(\la_j^+,\x_k^+) \re^\ph \prod_{l=1}^n
	   \frac{\sh(\la_j^+ - \la_l^+ - \h)\sh(\la_j^+ - \x_l^+ + \h)}
	        {\sh(\la_j^+ - \la_l^+ + \h)\sh(\la_j^+ - \x_l^+ - \h)}
		\epc
\end{align}
and $G(\la,\x)$ is the solution of the linear integral equation
(\ref{defg}).
\begin{equation}
     Z_n (\{\la^+\}|\{\x^+\}|\{\x^-\})
        = \prod_{j=1}^{m-n} \biggl[ 1 + \re^\ph \fa(\x_j^-)
	     \prod_{k=1}^n
	     \frac{f(\x_j^-,\x_k^+) f(\la_k^+,\x_j^-)}
	          {f(\x_k^+,\x_j^-) f(\x_j^-,\la_k^+)} \biggr] \epp
\end{equation}
\end{lemma}
\subsection{From sums to integrals}
In the final step of our calculation we would now like to express the
sums over partitions in (\ref{sumrep}) as multiple integral over
certain canonical paths. We shall apply a similar rationale as in
\cite{KMT99b,KMST02a}.

Consider a function $f(\om_1, \dots, \om_n): {\mathbb C}^n \rightarrow
{\mathbb C}$, symmetric in its arguments, equal to zero if any two
of its arguments agree, and analytic on and inside the simple $n$-fold
contour ${\cal C}^n$. The function $1 + \fa (\om)$ is meromorphic
inside the rectangle $Q$ (see section \ref{sec:auxfun}), where its
only zeros are the simple zeros located at those Bethe roots which
characterize the leading eigenvalue of the quantum transfer matrix. 
Hence, the only poles of the meromorphic function $1/(1 + \fa(\om))$
inside ${\cal C}$ are simple poles at the Bethe roots with residues
\begin{equation}
     \res \Bigl\{ \frac{1}{1 + \fa(\om)} \Bigr\}\Bigr|_{\om = \la_j}
        = \frac{1}{\fa' (\la_j)} \epc \qd j = 1, \dots, N/2 \epp
\end{equation}
It follows that
\begin{equation} \label{sumtoint}
     \frac{1}{n!} \int_{{\cal C}^n} \Bigl[ \prod_{j=1}^n 
        \frac{d \om_j}{2 \p \i (1 + \fa(\om_j))} \Bigr]
	f(\om_1, \dots, \om_n)
	= \mspace{-18mu}
	  \sum_{\substack{(\{\la^+\},\{\la^-\}) \in p_2 \{\la\} \\
	                  |\la^+| = n}}
	  \frac{f(\la_1^+, \dots, \la_n^+)}
	       {\prod_{j=1}^n \fa'(\la_j^+)} \epp
\end{equation}
Below we shall use this formula from the right to the left in order
to rewrite sums over partitions as multiple integrals.

In order to apply (\ref{sumtoint}) we split the sum in (\ref{sumrep})
as
\begin{equation} \label{splitsum}
     \sum_{\substack{(\{\la^+\},\{\la^-\}) \in p_2 \{\la\}\\
                     (\{\x^+\},\{\x^-\}) \in p_2 \{\x\}\\
		     |\x^+| + |\la^-| = N/2}}
        = \sum_{n=0}^m \:\:\:
	  \sum_{\substack{(\{\x^+\},\{\x^-\}) \in p_2 \{\x\} \\
	                  |\x^+| = n}} \:\:\:
	  \sum_{\substack{(\{\la^+\},\{\la^-\}) \in p_2 \{\la\} \\
	                  |\la^+| = n}} \epp
\end{equation}
We would like to apply (\ref{sumtoint}) to the rightmost sum in
(\ref{splitsum}) after inserting it into (\ref{sumrep}). However,
$Y_n (\{\la^+\}|\{\x^+\}) \, Z_n (\{\la^+\}|\{\x^+\}|\{\x^-\})$ is
not analytic in the variables $\la_j^+$ inside ${\cal C}^n$, but for
every $\la_j^+$ has simple poles at $\la_j^+ = \x_k^+$ for $k = 1,
\dots, n$ (the function $G(\la,\x)$ has a simple pole at $\la = \x$
(see the appendix), the simple poles of $\det \widetilde{M}
(\la_j^+,\x_k^+)$ at $\la_j^+ = \x_k^+$ are compensated by the simple
zeros of $\fbq (\la_j^+)$). Recall that the $\x_j$ are free parameters.
We choose them distinct from the Bethe roots $\la_j$ and all inside
${\cal C}$. Then there is a simple closed contour $\G$ encircling all
the $\x_j$ but none of the Bethe roots $\la_j$, and
$Y_n (\{\la^+\}|\{\x^+\}) \, Z_n (\{\la^+\}| \{\x^+\}| \{\x^-\})$ is
analytic inside ${\cal C} - \G$ (see figure
\begin{figure}

\begin{center}

\epsfxsize 9cm
\epsffile{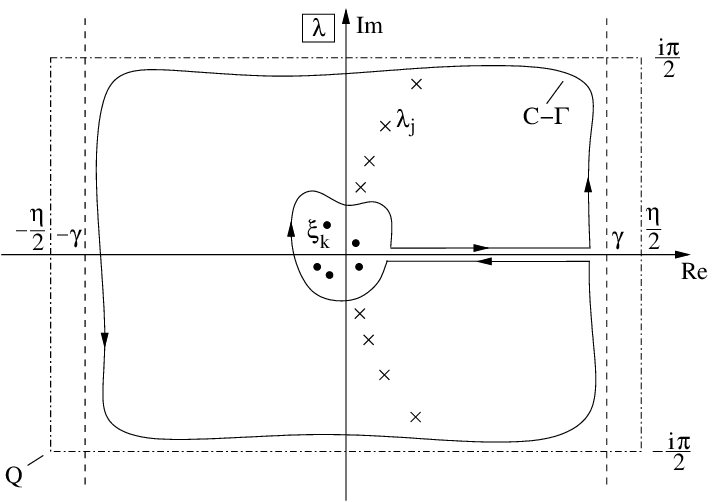}

\caption{\label{fig:cminusg} The path ${\cal C} - \G$.}
\end{center}

\end{figure}
\ref{fig:cminusg}). Thus, (\ref{sumtoint}) applies to ${\cal C} - \G$,
and we obtain
\begin{multline} \label{sumtointappl}
     \sum_{\substack{(\{\la^+\},\{\la^-\}) \in p_2 \{\la\} \\
                     |\la^+| = n}}
        \frac{Y_n (\{\la^+\}|\{\x^+\}) \,
	      Z_n (\{\la^+\}| \{\x^+\}| \{\x^-\})}
	     {\Bigl[ \prod_{j=1}^n \fa'(\la_j^+) \Bigr]
	      \Bigl[ \prod_{j=1}^{m-n} (1 + \fa(\x_j^-)) \Bigr]}\\[-1ex]
	= \frac{1}{n!} \int_{({\cal C} - \G)^n} \Bigl[ \prod_{j=1}^n 
	  \frac{d \om_j}{2 \p \i (1 + \fa(\om_j))} \Bigr]
	     \Bigl[ \prod_{j=1}^{m-n}
	  \frac{1}{(1 + \fa(\x_j^-))} \Bigr] \\
	     Y_n \bigl(\{\om_j\}_{j=1}^n|\{\x^+\}\bigr) \,
	     Z_n \bigl(\{\om_j\}_{j=1}^n| \{\x^+\}| \{\x^-\}\bigr) \epp
\end{multline}
This is already a multiple integral representation. Note, however, that
it is still inappropriate for performing the homogeneous limit and the
Trotter limit for two reasons. First, the integrand still contains
the auxiliary function $\fa (\x_j^-)$ which cannot be removed by
applying a formula like (\ref{sumtoint}) to the variables $\x_k$.
But the Trotter limit and the limit $\x \rightarrow 0$ in $\fa (\x)$
do not commute. Second, since the origin is a limit point of the
sequence of Bethe numbers that determine the leading eigenvalue of
the quantum transfer matrix in the Trotter limit, we could not avoid
that Bethe roots would cross the contour $\G$ if it would include the
origin which in turn would be necessary in order to perform the
homogeneous limit. The way out of these problems is amazingly
straightforward. In the next two sections we calculate the
$\G$-integrals explicitly and sum up the resulting terms. In this
procedure all unpleasant terms $\fa (\x_j^-)$ cancel each other.

\subsection{Evaluation of the $\pmb{\G}$-integrals}
Due to the symmetry in the $\om_j$ of the integrand in
(\ref{sumtointappl}) we may replace the integral with
\begin{equation} \label{intdecomp}
     \int_{({\cal C} - \G)^n} \prod_{j=1}^n d \om_j
        = \sum_{k=0}^n \binom{n}{k} (-1)^k
	  \int_{{\cal C}^{n-k}} \Bigl[ \prod_{j=1}^{n-k} d \om_j \Bigr]
	  \int_{\G^k} \Bigl[ \prod_{j=1}^k d \om_j \Bigr] \epp
\end{equation}
Our next task is to evaluate the $\G$-integrals. This can be achieved
by means of the residue theorem. We obtain the following result,
\begin{align} \label{gammaint}
     \int_{\G^k} \Bigl[ \prod_{j=1}^k &
        \frac{d \om_{n - k + j}}
	     {2 \p \i (1 + \fa(\om_{n - k + j}))} \Bigr]
	     Y_n \bigl(\{\om_j\}_{j=1}^n|\{\x^+\}\bigr) \,
	     Z_n \bigl(\{\om_j\}_{j=1}^n|\{\x^+\}|\{\x^-\}\bigr)
	     \notag \\ &
     = k! \sum_{\substack{(\{\x^{++}\},\{\x^{+-}\}) \in p_2 \{\x^+\} \\
                          |\x^{+-}| = k}}
	  Y_{n-k} \bigl(\{\om_j\}_{j=1}^{n-k}|\{\x^{++}\}\bigr) \,
	  \Bigl[ \prod_{j=1}^k \frac{1}{(1 + \fa(\x_j^{+-}))}
	     \Bigr] \notag \\[-2ex] & \mspace{180mu}
          \biggl[ \prod_{j=1}^{m-n} \Bigl( 1 + \re^\ph \fa (\x_j^-)
	          \prod_{l=1}^{n-k}
		  \frac{f(\x_j^-,\x_l^{++}) f(\om_l,\x_j^-)}	
		       {f(\x_l^{++},\x_j^-) f(\x_j^-,\om_l)} \Bigr)
		       \biggr] \notag \\ & \mspace{180mu}
          \biggl[ \prod_{j=1}^k \Bigl( 1 - \re^\ph
	          \prod_{l=1}^{n-k}
		  \frac{f(\x_j^{+-},\x_l^{++}) f(\om_l,\x_j^{+-})}	
		       {f(\x_l^{++},\x_j^{+-}) f(\x_j^{+-},\om_l)}
		       \Bigr) \biggr] \epp
\end{align}

\subsection{Resummation of the $\pmb{\cal C}$-integrals}
\label{sec:resum}
Inserting (\ref{sumtointappl})-(\ref{gammaint}) into (\ref{sumrep})
we obtain an expression that contains only integrals over the canonical
contours ${\cal C}$. Yet the summation with respect to the inhomogeneity
parameters $\x_j$ looks more complicated than before. It turns out that
these sums can be simplified considerably by applying the following
\begin{lemma}
For any function $F(\x)$ the identity
\begin{multline}
     \sum_{k=0}^{|\x|} (-1)^k \mspace{-18mu}
     \sum_{\substack{(\{\x^+\},\{\x^-\}) \in p_2 \{\x\} \\
                     |\x^+| = k}}
     \Bigl[ \prod_{j=1}^{|\x^-|}
            \bigl( 1 + \re^\ph \fa (\x_j^-) F(\x_j^-) \bigr) \Bigr]
     \Bigl[ \prod_{j=1}^{|\x^+|}
            \bigl( 1 - \re^\ph F(\x_j^+) \bigr) \Bigr] \\[-4ex]
     = \re^{|\x|\ph} \prod_{j=1}^{|\x|} F(\x_j) (1 + \fa (\x_j))
\end{multline}
holds.
\end{lemma}
Thanks to this lemma we end up with the equation
\begin{multline} \label{mireplambda}
     \PH_N (\ph|\{\x\}) = \sum_{n=0}^m \frac{\re^{(m-n)\ph}}{n!}
        \sum_{\substack{(\{\x^+\},\{\x^-\}) \in p_2 \{\x\} \\
                        |\x^+| = n}}
        \biggl[ \prod_{j=1}^n \frac{1}{\fb' (\x_j^+)} \biggr] \\
	\int_{{\cal C}^n} 
        \biggl[ \prod_{j=1}^n
	        \frac{d \om_j \; \fb (\om_j)}
		     {2 \p \i (1 + \fa (\om_j))} \biggr]
        \biggl[ \prod_{j,k=1}^n
	     \frac{\sh(\om_j - \x_k^+ - \h)}{\sh(\x_j^+ - \x_k^+ - \h)}
	          \biggr]
        \det M (\om_j,\x_k^+) \; \det G(\om_j,\x_k^+)
\end{multline}
for the finite Trotter number approximant to the generating function
(\ref{genfun}) of the $\s^z$-$\s^z$ correlation functions. Here we
introduced the new abbreviations
\begin{align}
     \fb (\la) & = \prod_{k=1}^m \frac{1}{f(\x_k,\la)} \epc \\
     M (\om_j,\x_k^+) & = t(\x_k^+,\om_j) \mspace{-2mu}
        \prod_{l=1}^n \frac{\sh(\om_j - \x_l^+ - \h)}
	                   {\sh(\om_j - \om_l - \h)}
	+ t(\om_j,\x_k^+) \re^\ph \prod_{l=1}^n \mspace{-3mu}
	  \frac{\sh(\om_j - \x_l^+ + \h)}{\sh(\om_j - \om_l + \h)}.
\end{align}

The integral on the right hand side of (\ref{mireplambda}) is analytic
and symmetric in the $\x_j^+$ inside ${\cal C}^n$. It vanishes if any
two of the $\x_j$ are identical. The function $1/\fb (\la)$ is
meromorphic inside ${\cal C}$ with only simple poles located at $\la
= \x_j$. Hence, we can apply (\ref{sumtoint}) to any simple closed
contour $\G$ which lies inside ${\cal C}$ and encircles all the
inhomogeneities $\x_j$. It follows that
\begin{multline} \label{mirepn}
     \PH_N (\ph|\{\x\}) = \sum_{n=0}^m \frac{\re^{(m-n)\ph}}{(n!)^2}
	\biggl[ \prod_{j=1}^n \int_{\G}
	   \frac{d \z_j}{2 \p \i \; \fb (\z_j)}
	\int_{\cal C} \frac{d \om_j \; \fb (\om_j)}
	                   {2 \p \i (1 + \fa (\om_j))} \biggr] \\
        \biggl[ \prod_{j,k=1}^n
	     \frac{\sh(\om_j - \z_k - \h)}{\sh(\z_j - \z_k - \h)}
	          \biggr]
        \det M (\om_j,\z_k) \; \det G(\om_j,\z_k) \epp
\end{multline}
This is our final formula for the finite Trotter number approximant
$\PH_N (\ph|\{\x\})$. We are now in a position to take the homogeneous
limit and the Trotter limit analytically.

\subsection{Taking the limits}
\label{sec:lim}
The information about the inhomogeneities is contained in the function
$\fb (\la)$. From its definition we see immediately that
\begin{equation}
     \lim_{\x_1, \dots, \x_m \rightarrow 0} \fb(\la)
        = \biggl( \frac{\sh(\la)}{\sh(\la - \h)}
	                \biggr)^{\mspace{-6mu} m} \epp
\end{equation}
The information about the discreteness of the Trotter decomposition,
on the other hand, is implicit in the function $\fa (\la)$. Taking
the Trotter limit means to take $\fa (\la)$ as a solution of equation
(\ref{nlieh}). Thus, we conclude and summarize our main result.
\begin{theorem}
\label{theorem:main}
The generating function of the $\s^z$-$\s^z$ correlation functions
of the XXZ chain has the following integral representation:
\begin{multline} \label{mirep}
     \bigl\< \exp\bigl\{ \ph \tst{\sum_{n=1}^m} {e_n}_2^2 \bigr\}
        \bigr\>_{T,h} \\[1ex]
     = \sum_{n=0}^m \frac{\re^{(m-n)\ph}}{(n!)^2}
	\biggl[ \prod_{j=1}^n \int_{\G}
	   \frac{d \z_j}{2 \p \i}
           \biggl( \frac{\sh(\z_j - \h)}{\sh(\z_j)}
	           \biggr)^{\mspace{-6mu} m}
	\int_{\cal C} \frac{d \om_j}{2 \p \i (1 + \fa (\om_j))}
        \biggl( \frac{\sh(\om_j)}{\sh(\om_j - \h)}
	              \biggr)^{\mspace{-6mu} m} \: \biggr] \\
        \biggl[ \prod_{j,k=1}^n
	     \frac{\sh(\om_j - \z_k - \h)}{\sh(\z_j - \z_k - \h)}
	          \biggr]
        \det M (\om_j,\z_k) \; \det G(\om_j,\z_k) \epc
\end{multline}
where $\fa (\la)$ is the auxiliary function (\ref{nlieh}) that
determines the free energy in the Trotter limit,
\begin{equation}
     M (\om_j,\z_k) = t(\z_k,\om_j)
        \prod_{l=1}^n \frac{\sh(\om_j - \z_l - \h)}
	                   {\sh(\om_j - \om_l - \h)}
	+ t(\om_j,\z_k) \re^\ph \prod_{l=1}^n
	  \frac{\sh(\om_j - \z_l + \h)}{\sh(\om_j - \om_l + \h)} \epc
\end{equation}
and the function $G(\la,\z)$ is the solution of the linear integral
equation
\begin{equation} \label{repdefg}
     G(\la,\z) = t(\z,\la)
                  + \int_{\cal C} \frac{d \om}{2 \p \i} \,
	            \frac{\sh (2 \h)}
		         {\sh(\la - \om + \h)\sh(\la - \om - \h)}
	            \frac{G (\om,\z)}{1 + \fa (\om)} \epp
\end{equation}
The contour ${\cal C}$ is the canonical contour in the non-linear
integral equation formalism (see figure \ref{fig:cancon}). $\G$ is
any simple closed contour which lies inside ${\cal C}$ and encircles
the origin.
\end{theorem}

As we have seen in sections \ref{sec:magfie}, \ref{sec:freemandela}
both auxiliary function, $\fa$ and $\faq$, can be equivalently used
to express physical quantities. To switch from one representation to
the other one can use the identity
\begin{equation} \label{switchaaq}
     \frac{1}{1 + \fa (\la)} = 1 - \frac{1}{1 + \faq (\la)} \epp
\end{equation}
Inserting it into (\ref{repdefg}) and using
\begin{equation}
     \frac{\sh (2 \h)}{\sh(\la - \z + \h)\sh(\la - \z - \h)}
        = t(\la,\z) + t(\z,\la)
\end{equation}
we find, for instance,
\begin{equation} \label{repdefgq}
     G(\la,\z) = - t(\la,\z)
                  - \int_{\cal C} \frac{d \om}{2 \p \i} \,
	            \frac{\sh (2 \h)}
		         {\sh(\la - \om + \h)\sh(\la - \om - \h)}
	            \frac{G (\om,\z)}{1 + \faq (\om)}
\end{equation}
which equally well defines the inhomogeneous density function
$G(\la,\z)$. The task of re-expressing the multiple integral
representation (\ref{mirep}) in terms of $\faq$ is, of course, more
cumbersome. One may, for instance, start from (\ref{mirepn}), insert
(\ref{switchaaq}), evaluate the integrals which do not contain
$\faq$, and resum the resulting terms in a similar way as in
section \ref{sec:resum}. One arrives at the following
\begin{corollary}
When expressed in terms of $\faq$ the generating function takes the
form
\begin{multline} \label{mirepq}
     \bigl\< \exp\bigl\{ \ph \tst{\sum_{n=1}^m} {e_n}_2^2 \bigr\}
        \bigr\>_{T,h} \\[1ex]
     = \sum_{n=0}^m \frac{(-1)^n}{(n!)^2}
	\biggl[ \prod_{j=1}^n \int_{\G}
	   \frac{d \z_j}{2 \p \i}
           \biggl( \frac{\sh(\z_j + \h)}{\sh(\z_j)}
	           \biggr)^{\mspace{-6mu} m}
	\int_{\cal C} \frac{d \om_j}{2 \p \i (1 + \faq (\om_j))}
        \biggl( \frac{\sh(\om_j)}{\sh(\om_j + \h)}
	              \biggr)^{\mspace{-6mu} m} \: \biggr] \\
        \biggl[ \prod_{j,k=1}^n
	     \frac{\sh(\om_j - \z_k + \h)}{\sh(\z_j - \z_k + \h)}
	          \biggr]
        \det M (\om_j,\z_k) \; \det G(\om_j,\z_k) \epc
\end{multline}
where $M (\om_j,\z_k)$ and $G(\om_j,\z_k)$ are the same as in theorem
\ref{theorem:main}.
\end{corollary}
In the remaining sections we shall discuss several special cases and
limits of (\ref{mirep}) and (\ref{mirepq}).

\subsection{One-point function}
We begin with the one-point function obtained from (\ref{mirep}) or
(\ref{mirepq}), respectively, for $m=1$. In this case the sums on the
right hand side of (\ref{mirep}) and (\ref{mirepq}) consist of only
two terms, the first of which does not contain any integral. In the
second terms the integration over $\G$ can easily be performed, and
one remains with the formulae
\begin{equation} \label{onepoint}
     \bigl\< \bigl(
       \begin{smallmatrix} 1 & 0 \\ 0 & e^\ph \end{smallmatrix}
       \bigr)_1  \bigr\>_{T,h}
       = \re^\ph - (1 - \re^\ph)
                   \int_{\cal C} \frac{d \om}{2 \p \i}
		   \frac{G(\om)}{1 + \fa (\om)}
       = 1 + (1 - \re^\ph) \int_{\cal C} \frac{d \om}{2 \p \i}
                           \frac{G(\om)}{1 + \faq (\om)} \epp
\end{equation}
Here we inserted the identity $G(\om) = G(\om,0)$ which is obtained
by comparing (\ref{defghom}) and (\ref{repdefg}).

Setting $\ph = 0$ in the above equation both expressions for the
one-point function yield 1 as it has to be. Setting $\ph = \i \p$,
on the other hand, and multiplying (\ref{onepoint}) by $\2$ we obtain
two formulae for the magnetization,
\begin{equation} \label{szexp}
     \bigl\< S^z_1  \bigr\>_{T,h}
       = - \2  - \int_{\cal C} \frac{d \om}{2 \p \i}
                               \frac{G(\om)}{1 + \fa (\om)}
       = \2 + \int_{\cal C} \frac{d \om}{2 \p \i}
                            \frac{G(\om)}{1 + \faq (\om)} \epp
\end{equation}
These formulae are the same as obtained in section \ref{sec:freemandela}
based on a completely different reasoning, namely by taking the
derivative of the free energy with respect to the magnetic field.

\subsection{Emptiness formation probability}
In the limiting cases $\ph \rightarrow \pm \infty$ (for $\ph
\rightarrow \infty$ one has to multiply by $\re^{- m \ph}$ first in
order to obtain a finite result) our formulae (\ref{mirep}),
(\ref{mirepq}) simplify considerably. The resulting expressions
on the left hand side of (\ref{mirep}), (\ref{mirepq}),
$\<{e_1}_1^1 \dots {e_m}_1^1\>_{T,h}$ or $\<{e_1}_2^2 \dots
{e_m}_2^2\>_{T,h}$, respectively, describe the probability to find
a string of up- or down-spins of length~$m$. For spin chains these
probabilities where introduced in \cite{KIEU94}. They come under
the name of the `emptiness formation probability' and recently became
popular, since the integrals in the zero temperature multiple integral
representation of the emptiness formation probability \cite{JMMN92,%
KIEU94} could be evaluated explicitly for small $m$ \cite{BoKo01}.

For finite temperatures our formulae (\ref{mirep}) and (\ref{mirepq})
yield the two alternative expressions
\begin{subequations}
\begin{align}
     \bigl\< {e_1}_1^1 \dots & {e_m}_1^1 \bigr\>_{T,h} \notag \\[1ex]
     = & \frac{1}{(m!)^2}
	\biggl[ \prod_{j=1}^m \int_{\G}
	   \frac{d \z_j}{2 \p \i}
           \biggl( \frac{\sh(\z_j - \h)}{\sh(\z_j)}
	           \biggr)^{\mspace{-6mu} m}
	\int_{\cal C} \frac{d \om_j}{2 \p \i (1 + \fa (\om_j))}
        \biggl( \frac{\sh(\om_j)}{\sh(\om_j - \h)}
	              \biggr)^{\mspace{-6mu} m} \: \biggr] \notag \\ &
        \biggl[ \prod_{j,k=1}^n
	     \frac{\sh^2 (\om_j - \z_k - \h)}
	          {\sh(\z_j - \z_k - \h) \sh(\om_j - \om_k - \h)}
	          \biggr]
        \det \, t (\z_k,\om_j) \; \det G(\om_j,\z_k)
	\displaybreak[0] \\[1ex]
     = & \sum_{n=0}^m \frac{(-1)^n}{(n!)^2}
	\biggl[ \prod_{j=1}^n \int_{\G}
	   \frac{d \z_j}{2 \p \i}
           \biggl( \frac{\sh(\z_j + \h)}{\sh(\z_j)}
	           \biggr)^{\mspace{-6mu} m}
	\int_{\cal C} \frac{d \om_j}{2 \p \i (1 + \faq (\om_j))}
        \biggl( \frac{\sh(\om_j)}{\sh(\om_j + \h)}
	              \biggr)^{\mspace{-6mu} m} \: \biggr] \notag \\ &
        \biggl[ \prod_{j,k=1}^n
	     \frac{\sh(\om_j - \z_k + \h)\sh(\om_j - \z_k - \h)}
	          {\sh(\z_j - \z_k + \h)\sh(\om_j - \om_k - \h)}
	          \biggr]
        \det \, t (\z_k,\om_j) \; \det G(\om_j,\z_k)
\end{align}
\end{subequations}
for $\bigl\< {e_1}_1^1 \dots {e_m}_1^1 \bigr\>_{T,h}$ and, similarly,
\begin{subequations}
\begin{align}
     \bigl\< {e_1}_2^2 \dots & {e_m}_2^2 \bigr\>_{T,h} \notag \\[1ex]
     = & \frac{(-1)^m}{(m!)^2}
	\biggl[ \prod_{j=1}^m \int_{\G}
	   \frac{d \z_j}{2 \p \i}
           \biggl( \frac{\sh(\z_j + \h)}{\sh(\z_j)}
	           \biggr)^{\mspace{-6mu} m}
	\int_{\cal C} \frac{d \om_j}{2 \p \i (1 + \faq (\om_j))}
        \biggl( \frac{\sh(\om_j)}{\sh(\om_j + \h)}
	              \biggr)^{\mspace{-6mu} m} \: \biggr] \notag \\ &
        \biggl[ \prod_{j,k=1}^n
	     \frac{\sh^2 (\om_j - \z_k + \h)}
	          {\sh(\z_j - \z_k + \h) \sh(\om_j - \om_k + \h)}
	          \biggr]
        \det \, t (\om_j,\z_k) \; \det G(\om_j,\z_k) \\[1ex]
     = & \sum_{n=0}^m \frac{1}{(n!)^2}
	\biggl[ \prod_{j=1}^n \int_{\G}
	   \frac{d \z_j}{2 \p \i}
           \biggl( \frac{\sh(\z_j - \h)}{\sh(\z_j)}
	           \biggr)^{\mspace{-6mu} m}
	\int_{\cal C} \frac{d \om_j}{2 \p \i (1 + \fa (\om_j))}
        \biggl( \frac{\sh(\om_j)}{\sh(\om_j - \h)}
	              \biggr)^{\mspace{-6mu} m} \: \biggr] \notag \\ &
        \biggl[ \prod_{j,k=1}^n
	     \frac{\sh(\om_j - \z_k + \h)\sh(\om_j - \z_k - \h)}
	          {\sh(\z_j - \z_k + \h)\sh(\om_j - \om_k - \h)}
	          \biggr]
        \det \, t (\om_j,\z_k) \; \det G(\om_j,\z_k) \epp
\end{align}
\end{subequations}

\subsection{Zero temperature limit}
How can we make contact with the zero temperature results obtained in
\cite{KMST02a}? It is clear by inspection that the auxiliary functions
$\fa$ and $\faq$ defined by (\ref{nlieh}) and (\ref{nliehaq}) do not
approach a limit as $T$ goes to zero. The inhomogeneities are unbounded
in $T$. In order to obtain a sensible result we have to multiply by $T$
first. Setting $\e (\la) = - T \ln (\fa (\la)) = T \ln (\faq (\la))$
the non-linear integral equations turn into
\begin{subequations} \label{nlieheps}
\begin{align} \label{epsa}
     \e (\la) & = h + \frac{2J \sh^2 (\h)}{\sh(\la) \sh(\la + \h)}
                  + \int_{\cal C} \frac{d \om}{2 \p \i} \,
	            \frac{\sh (2 \h) T \ln (1 + \re^{- \e (\om)/T})}
		         {\sh(\la - \om + \h) \sh(\la - \om - \h)} \\
              & = h - \frac{2J \sh^2 (\h)}{\sh(\la) \sh(\la - \h)}
                  + \int_{\cal C} \frac{d \om}{2 \p \i} \,
	            \frac{\sh (2 \h) T \ln (1 + \re^{\e (\om)/T})}
		         {\sh(\la - \om + \h) \sh(\la - \om - \h)} \epp
			 \label{epsb}
\end{align}
\end{subequations}

Here we shift the left and right edges of the integration contour to
$\pm \h/2$, where $\e (\la)$ is real (recall that the contributions
from the upper and the lower edge cancel each other). We see that it
only matters if $\e (\la)$ is positive or negative as $T$ goes to zero.
In (\ref{epsa}) only the negative part of $\e (\la)$ on the contour
contributes to the zero temperature limit, in (\ref{epsb}) only the
positive part. This divides the contour ${\cal C}$ into two disjoint
pieces ${\cal C}^{(+)}$ and ${\cal C}^{(-)}$ on which $\e (\la)$ can be
calculated by means of the linear integral equations
\begin{subequations} \label{epspm}
\begin{align} \label{epsminus}
     \e (\la) & = h + \frac{2J \sh^2 (\h)}{\sh(\la) \sh(\la + \h)}
                  - \int_{{\cal C}^{(-)}} \frac{d \om}{2 \p \i} \,
	            \frac{\sh (2 \h) \e (\om)}
		         {\sh(\la - \om + \h) \sh(\la - \om - \h)}
			 \epc \\
     \e (\la) & = h - \frac{2J \sh^2 (\h)}{\sh(\la) \sh(\la - \h)}
                  + \int_{{\cal C}^{(+)}} \frac{d \om}{2 \p \i} \,
	            \frac{\sh (2 \h) \e (\om)}
		         {\sh(\la - \om + \h) \sh(\la - \om - \h)} \epp
			 \label{epsplus}
\end{align}
\end{subequations}
These integral equations can be transformed into integral equations
for dressed energies, which is, however, not our actual aim. The most
important conclusion instead is that
\begin{equation} \label{freeze}
     \frac{1}{1 + \faq (\la)} = \frac{1}{1 + \re^{\e (\la)/T}}
        \qd \xrightarrow{T \rightarrow 0} \qd
	\begin{cases} 0 & \text{on ${\cal C}^{(+)}$} \\
	              1 & \text{on ${\cal C}^{(-)}$} \end{cases} \epp
\end{equation}
This function behaves like a `Fermi function on $\cal C$'.

Applying (\ref{freeze}) to the integral equation (\ref{repdefgq}) and
shifting the arguments appropriately we obtain
\begin{equation} \label{repdefgqzero}
     - G \bigl(\la - \tst{\frac{\h}{2}},\z - \tst{\frac{\h}{2}}\bigr)
        - \int_{{\cal C}^{(-)} + \frac{\h}{2}} \frac{d \om}{2 \p \i} \,
          \frac{\sh (2 \h) \, G \bigl(\om - \tst{\frac{\h}{2}},
	                        \z - \tst{\frac{\h}{2}}\bigr)}
	       {\sh(\la - \om + \h)\sh(\la - \om - \h)} = t(\la,\z)
	       \epp
\end{equation}
This equation has to be compared with equation (2.16) of \cite{KMST02a}.
The two equations agree if we set
\begin{equation} \label{defrho}
     G \bigl(\la - \tst{\frac{\h}{2}},\z - \tst{\frac{\h}{2}}\bigr)
        = 2 \p \i \, \r (\la, \z)
\end{equation}
and identify the integration contour in equation (2.16) of
\cite{KMST02a} with $- ({\cal C}^{(-)} + \frac{\h}{2})$. A close
inspection (for details see e.g.\ \cite{KlSc03}) of (\ref{nlieheps}),
(\ref{epspm}) shows indeed that (for $h > 0$, $\h < 0$!) the contour
${\cal C}^{(-)}$ is a line segment $\Re \la = \h/2$, $- \a < \Im \la
< \a$ (where $\a > 0$ depends on the magnetic field $h$). This is
sufficient to identify (\ref{repdefgqzero}) with (2.16) of
\cite{KMST02a}.

Let us introduce the shorthand notation ${\cal I} = - ({\cal C}^{(-)} +
\frac{\h}{2})$. Then, using (\ref{freeze}) and (\ref{defrho}), our
second expression (\ref{mirepq}) for the generating function of the
$\s^z$-$\s^z$ correlation functions can be evaluated in the zero
temperature limit. We obtain
\begin{multline} \label{mirepqtzero}
     \lim_{T \rightarrow 0}
     \bigl\< \exp\bigl\{ \ph \tst{\sum_{n=1}^m} {e_n}_2^2 \bigr\}
        \bigr\>_{T,h} \\[1ex]
     = \sum_{n=0}^m \frac{1}{(n!)^2}
	\biggl[ \prod_{j=1}^n \int_{\G + \frac{\h}{2}}
	   \frac{d \z_j}{2 \p \i}
           \biggl( \frac{\sh(\z_j + \frac{\h}{2})}
	                {\sh(\z_j - \frac{\h}{2})}
	           \biggr)^{\mspace{-6mu} m}
	\int_{\cal I} d \om_j
        \biggl( \frac{\sh(\om_j - \frac{\h}{2})}
	             {\sh(\om_j + \frac{\h}{2})}
	              \biggr)^{\mspace{-6mu} m} \: \biggr] \\
        \biggl[ \prod_{j,k=1}^n
	     \frac{\sh(\om_j - \z_k + \h)}{\sh(\z_j - \z_k + \h)}
	          \biggr]
        \det M (\om_j,\z_k) \; \det \r(\om_j,\z_k) \epc
\end{multline}
where $\r (\la, \z)$ is the solution of the linear integral equation
\begin{equation} \label{lierho}
     - 2 \p \i \, \r (\la, \z) + \int_{\cal I} d \om \,
          \frac{\sh (2 \h) \, \r (\om, \z)}
	       {\sh(\la - \om + \h)\sh(\la - \om - \h)} = t(\la,\z)
\end{equation}
which is precisely the result of \cite{KMST02a}.

\subsection{High temperature limit}
In the limit of infinite temperature all correlations should vanish.
How can this be seen from our formulae? For $T \rightarrow \infty$
the inhomogeneity on the right hand side of the integral equations
(\ref{nlieh}) and (\ref{nliehaq}) vanishes. The remaining homogeneous
equations are solved by $\fa (\la) = 1$ or $\faq (\la) = 1$,
respectively. Inserting this, for instance, into (\ref{mirep}) we see
that the only singularities in the resulting integrand are the
simple poles of the function $G(\om,\z)$. Therefore the integrals
can be calculated by means of the residue theorem. Using similar
techniques as described in section \ref{sec:resum} we can simplify
the result which finally turns into
\begin{equation}
     \lim_{T \rightarrow \infty}
     \bigl\< \exp\bigl\{ \ph \tst{\sum_{n=1}^m} {e_n}_2^2 \bigr\}
        \bigr\>_{T,h} = \biggl( \frac{1 + \re^\ph}{2}
	                                \biggr)^{\mspace{-6mu} m} \epp
\end{equation}
Applying (\ref{genfunappl}) we obtain
\begin{equation}
     \lim_{T \rightarrow \infty}
     \bigl\< \s_1^z \s_m^z \bigr\>_{T,h} = 0
\end{equation}
as we expected.

\section{Conclusions} \label{sec:concl}
In this work integral representations for finite temperature
correlation functions of an integrable model have been derived for the
first time. We concentrated on the generating function (\ref{genfun})
of the $\s^z$-$\s^z$ correlation functions, since it appeared
particularly appropriate to us for developing the method. Now this
method may be applied to other generating functions and also directly
to two-point functions. Work in this direction is on the way.

It will be very interesting to further study the integral
representations. How can they be utilized to obtain the long-distance
asymptotics? How efficient are they for numerical calculations? Are
there special cases where part of the integrations can be performed
analytically? We hope to address these questions in forthcoming
publications.

\begin{figure}
\begin{center}

\epsfxsize 9cm
\epsffile{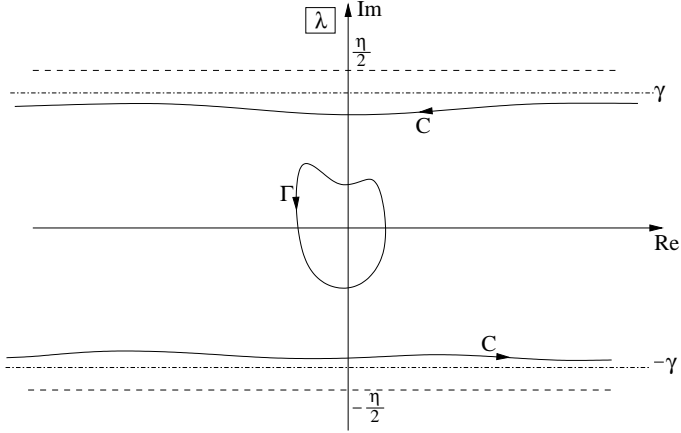}

\caption{\label{fig:critcont} Contours of integration in the critical
regime for $0 \le \D < 1$. For $- 1 < \D \le 0$ one has to choose
$|\g| < \frac{\p}{2} - \frac{|\h|}{2}$. Alternatively one may switch
the sign of $J$ making use of the equivalence of the parameter
regimes $- 1 < \D \le 0$, $J > 0$ and $0 \le \D < 1$, $J < 0$.}
\end{center}

\end{figure}

We restricted our presentation to the off-critical regime $\D > 1$.
We would like to stress, however, that our results straightforwardly
extend to the critical regime $|\D| \le 1$, where only the canonical
contour has to be redefined (see figure \ref{fig:critcont}). 
The situation is much the same as in the zero temperature limit
\cite{KMST02a}.

{\bf Acknowledgement.}
The authors would like to thank M. Karbach for helpful discussions.

\clearpage

{\appendix
\Appendix{A ratio of determinants}
\noindent
Here we prove lemma \ref{lem:detrat}. We define a column vector
$\nv (\x)$ with entries
\begin{equation}
     \nv_j (\x) = t(\tilde \la_j, \x) - t(\x, \tilde \la_j) \fa (\x)
                  \epc \qd j = 1, \dots, N/2
\end{equation}
and another column vector
\begin{equation}
     \jv (\x) = N_0^{-1} \nv (\x) \epp
\end{equation}
The matrices $N_0$ and $N_n$ have the last $N/2 - n$ columns in common.
Thus, we conclude that
\begin{equation} \label{intj}
     \frac{\det N_n}{\det N_0} = \det N_0^{-1} N_n
        = \det (\jv (\tilde \x_1), \dots, \jv (\tilde \x_n), \ev_{n+1},
	        \dots, \ev_{N/2})
        = \det \, \< \ev_j, \jv (\tilde \x_k) \> \epc
\end{equation}
where the $\ev_j$ are the canonical unit column vectors having a single
non-zero entry $1$ in the $j$th row, and $j$ and $k$ on the right hand
side run from $1$ to $n$. In (\ref{intj}) we have expressed the ratio
of two $N/2 \times N/2$ determinants as a single $n \times n$
determinant, a trick we borrowed from \cite{KMT99b}.

In the second step of our proof we transform the equation
$N_0 \, \jv(\x) = \nv (\x)$ that determines $\jv$ into an integral
equation. For this purpose we first of all rewrite it in components,
\begin{equation}
     - \fa' (\tilde \la_k) \, \jv_k (\x)
     + \sum_{l=1}^{N/2} \frac{\sh(2\h) \, \jv_l (\x)}
                             {\sh(\tilde \la_k - \tilde \la_l + \h)
                              \sh(\tilde \la_k - \tilde \la_l - \h)}
     = t(\tilde \la_k, \x) - t(\x, \tilde \la_k) \fa (\x) \epp
\end{equation}
We define a function
\begin{equation} \label{deff}
     F(\la,\x) = t(\x, \la) \fa (\x) - t(\la, \x)
     + \sum_{l=1}^{N/2} \frac{\sh(2\h) \, \jv_l (\x)}
                             {\sh(\la - \tilde \la_l + \h)
                              \sh(\la - \tilde \la_l - \h)}
\end{equation}
and observe that this function determines $\jv (\x)$ through
\begin{equation}
     F(\tilde \la_k, \x) = \fa' (\tilde \la_k) \, \jv_k (\x) \epp
\end{equation}
We shall now assume that $\x$ is located inside our canonical contour
${\cal C}$ sketched in figure~\ref{fig:cancon}. Then the only
singularity of $F(\la,\x)$ as a function of $\la$ inside the rectangle
$Q$ (see again figure \ref{fig:cancon}) is a simple pole at $\la = \x$.
The residue at this pole is
\begin{equation} \label{resf}
     \res\{ F(\la, \x) \} \bigr|_{\la = \x}
        = \lim_{\la \rightarrow \x} \sh(\la - \x) F(\la, \x)
	= - 1 - \fa (\x) \epp
\end{equation}
This residue gives an additional contribution when we apply equation
(\ref{sumtoint}) to (\ref{deff}) in order to transform the sum on the
right hand side into an integral. We end up with
\begin{equation} \label{fviaint}
     F(\la,\x) = t(\x,\la)(1 + \fa (\x))
                  + \int_{\cal C} \frac{d \om}{2 \p \i} \,
	            \frac{\sh (2 \h)}
		         {\sh(\la - \om + \h)\sh(\la - \om - \h)}
	            \frac{F (\om,\x)}{1 + \fa (\om)} \epp
\end{equation}
The latter equation suggest to introduce a function
\begin{equation} \label{intg}
     G(\la,\x) = \frac{F(\la, \x)}{1 + \fa (\x)} \epp
\end{equation}
This function then solves the linear integral equation (\ref{defg})
defined in lemma \ref{lem:detrat}. Coming back to our original
notation (see (\ref{doptnot})) we may finally express the ratio of the
two determinants as
\begin{equation}
          \frac{\det N_n}{\det N_0}
	     = \biggl[ \prod_{l=1}^n
	       \frac{1 + \fa(\x_l^+)}{\fa'(\la_l^+)} \biggr]
	       \det G(\la_j^+,\x_k^+) \epc
\end{equation}
and our lemma is proven.

Note that is follows from (\ref{resf}) and (\ref{intg}) that the only
singularity of $G(\la,\x)$ (considered as a function of $\la$) inside
the rectangle $Q$ is a simple pole with residue $-1$ at $\la = \x$.
This fact is frequently used in the derivation of the multiple integral
representations (\ref{mirep}) and (\ref{mirepq}).
}


\end{document}